\documentclass[11pt,a4paper]{article}
\pdfoutput=1
\usepackage{jheppub}

\usepackage[english]{babel}
\usepackage{amsfonts}
\usepackage{graphicx}
\usepackage{amsmath}

\newcommand{\be}{\begin{equation}}
\newcommand{\ee}{\end{equation}}
\newcommand{\bea}{\begin{eqnarray}}
\newcommand{\eea}{\end{eqnarray}}

\begin{document}

%
%
%

\title{   Phase Transitions  of  Quintessential  AdS Black Holes in M-theory/Superstring  Inspired Models }
\author[a]{   A. Belhaj} 
\author[a]{A. El Balali} 
\author[a]{ W. El Hadri} 
\author[a]{ Y. Hassouni}
\author[b,c]{ E. Torrente-Lujan}

\affiliation[a]{
ESMAR, D\'{e}partement de Physique, Facult\'e des Sciences, Universit\'e Mohammed V de Rabat, \\    Rabat, Morocco} 
\affiliation[b]{ Fisica T\'eorica, Dep. de F\'isica, Univ.  de Murcia,
Campus de Espinardo, E-30100 Murcia, Spain} 
\affiliation[c]{TH-division, CERN, Geneve}

\emailAdd{adil.belhaj@um5.ac.ma}
\emailAdd{torrente@cern.ch}
\preprint{
\vspace{-0.8cm}
{\flushright{
FISPAC-TH/20-31415,  \\
UQBAR-TH/20-27182.\\
}}
}

\abstract{
We 
study 
$d$-dimensional $AdS$ black holes 
surrounded by Dark Energy (DE), 
embedded in $D$-dimensional M-theory/superstring 
inspired models having $AdS_d \times \mathbb{S}^{d+k}$ space-time  
with   $D=2d+k$. We focus on   the thermodynamic Hawking-Page phase 
transitions of quintessential DE 
black hole solutions,  whose microscopical origin 
is linked 
to   $N$ coincident $(d-2)$-branes supposed 
to live  in  such $(2d+k)$-dimensional models.
 Interpreting the cosmological constant as 
the number of colors $\propto N^{\frac{d-1}{2}}$, we calculate  various thermodynamical quantities 
in terms of brane number,  entropy and  DE contributions.  
Computing the chemical potential conjugated to the number of colors in the absence of DE, we  show  that a generic 
 black hole is more stable for  a larger number of branes 
for lower dimensions $d$.
In the presence of DE, we find   that the DE state parameter $\omega_q$ should take    particular values, for $(D,d,k)$ models,   providing a non trivial phase  transition structure.


}
%

\maketitle
\section{Introduction}

Recent observations, consistent with expectations for the shadow of a Kerr black hole (BH) as predicted
by general relativity, have been, for the first time, presented 
\cite{Akiyama:2019cqa,Akiyama:2019eap}. This result  emphasise the need to address the
long-standing puzzles presented by its BH solutions. A classical stationary BH solution is characterised
by its mass $M$, angular momentum $J$ and charge $Q$ alone. In particular, its horizon area is a
simple function of these three quantities. Identifying the horizon area as an entropy  the BHs obeys a set of laws  being 
directly analogous to those of thermodynamics \cite{B1,B2,BCH,H1,H2}. 
According to Hawking’s prediction \cite{H1,H2},  BHs 
 emit thermal radiation at the semiclassical level fixing the Bekenstein-Hawking
area/entropy relation to be  (where  $c = \hbar = G = 1$),
\begin{equation}
S_{BH} \sim \frac{A}{4}.
\end{equation}
This  suggests that the thermodynamic interpretation of BH mechanics is more than a mere analogy and points to the existence of sublying degrees of freedom\cite{SV}.
Any complete theory of quantum gravity should address this challenge in 
some way or at least advance in this direction.  More detailed view of  black holes could be found in \cite{I1,I2,I3,anas1,anas11,anas12,anas13,anas14,I4,I5,Adil,anas2,I15}.

The thermodynamic properties of $AdS$ black holes have been extensively 
studied, the existence of a  minimal Hawking temperature and  
the Hawking-Page phase transition \cite{I6,w1,w2,w3}. 
The Hawking-Page phase transition between large stable black holes and thermal
 gas in the AdS space  has been approached  using different methods  \cite{I7,I8}. 

For example,  an analogy between phase structures of various $AdS$ black holes
 and statistical modes associated with Van der Waals like phase transitions  
has been suggested\cite{anas3,I9,I10}.
 Interpreting the cosmological constant   as a kind of  thermodynamic pressure,
and its conjugate  variable as the thermodynamic volume,   several  
non trivial results have been presented  \cite{w4,I11,I12,I13,I14,I16,I17,I18,X0,Mth1,Mth2}.
Thermodynamics of  $AdS$ black holes, in supergravity theories,  have been
also investigated by exploiting  the  AdS/CFT correspondence  which provides an
 interplay  between gravitational models in $d$-dimensional $AdS$ geometries 
and $(d-1)$-dimensional conformal field theories living in the  boundary of 
such $AdS$ spaces.  Using  the physics  of solitonic  branes,  different models in 
type IIB superstrings and M-theory have been studied by considering the
 cosmological constant in the bulk closely related to the number of colors 
associated with branes in question. The thermodynamic stability behavior of such 
$AdS$ black holes in higher dimensional known supergravity theories has 
been examined in this context \cite{I16,I17,I18,X0,Mth1,Mth2}.

Dark Energy (DE) is needed to explain
  the, well established,  observation of the existence of an 
accelerated expansion of the universe.
On lack of a deeper microscopical explanation, the   
 ratio  of a  pressure and energy density 
$\omega_q=\frac{p_{dark}}{\rho_{dark}}$,  interpreted  as a DE fluid 
equation of state appearing in the Einstein stress tensor,  is usually 
used to model DE.   
Distinguished  DE models  have been discussed in  terms of   such  ratio
covering  the range $\left]-1,0\right[$ \cite{DE1}.  Among others, 
 "quintessence models",  associated with ratio values in 
the range $\left[-1,-\frac{1}{3}\right]$, interpreted as a dynamical field 
with a negative pressure  has been proposed in order to explain the 
universe acceleration \cite{I19,I21,I22}. 
In higher dimensional   theories, like M-theory supergravity,  
a  massless pseudoscalar axion like field, obtained from  the 
compactification to lower dimensions, has been considered   as a 
candidate to explain DE  contributions \cite{I20}.

Black holes could carry information about the nature of 
the elusive dark energy (DE), {\it et vice versa}. For example several BH 
solutions surrounded by a static spherically symmetric quintessence DE have 
been considered. Typically,  the presence of  DE acts as   a cooling fluid agent, 
largely  modifying several thermodynamical quantities   \cite{I23,I24, notrea}.

The aim of this work is to advance in the investigation of 
the thermodynamical phase transitions of $d$-dimensional $AdS$ black holes ($d \geq 4$) 
surrounded by quintessential DE described by the ratio $\omega_q$. 
These quintessential black holes solutions are  embedded in
 $D$-dimensional superstring/M-theory inspired models having 
$AdS_d \times \mathbb{S}^{d+k}$ space-time, where $D=2d+k$.   
These solutions, which  could be associated with $N$ coincident $(d-2)$-branes assumed to 
move in such higher dimensional models, are labeled  by a triplet $(D,d,k)$ where $k$  is associated with  
the internal space, the  $ \mathbb{S}^{d+k}$ sphere.   
By interpreting the cosmological constant as the number of colors (in fact 
proportional to $N^{\frac{d-1}{2}}$), we compute various thermodynamical quantities 
in terms of the brane number $N$, the entropy $S$ and  DE contributions. By computing the chemical 
potential conjugated to the number of colors in the absence of DE, we find that the black hole is more stable  for
  for configuration  with a larger number of branes, for small  dimensions $d$.
In the presence of DE, we observe that the state parameter $\omega_q$  takes 
 specific values, for $(D,d,k)$ models. Non trivial properties of the 
Hawking-Page  phase  transition in each case are obtained. 
In this work, we use dimensionless units in which one has $\hbar = c = 1$.


The organization of this paper is as follows. In section 2,   we provide detailed formulas for  
  $d$-dimensional AdS black holes embedded in $D$-dimensional superstring/M-theory inspired models  and 
compute several  thermodynamical quantities.  In sections 3 and 4, we study  in detail a model indexed   
by the triplet $(D,d,k)=(11,7,-3)$, associated with  the  compactification of M-theory on the 
sphere $\mathbb{S}^4$  with the  $M5$-branes without and with DE respectively. Similar results are presented 
in full detail in appendices.
In sections 5,   we present further discussions,   conclusions and open questions. The last sections
are devoted to appendices.  
In Appendices A, B, C and D,   we present  detailed 
results for $AdS_ 4\times \mathbb{S}^{7}$ and $AdS_ 5\times \mathbb{S}^{5}$  models  with
 and without DE.

\section{AdS black holes in M-theory/superstring inspired models}
\label{general}
In this work, we focus on the investigation of   $d$-dimensional AdS black holes embedded in $D$-dimensional superstring/M-theory inspired models (where $d \geq 4$).


We assume they are obtained from the compactification on $(D-d)$-dimensional real spheres 
denoted by $ \mathbb{S}^{D-d}$. In the presence of brane solitonic objects, the associated $D$-dimensional 
geometry can be factorized as follows
\begin{equation}
AdS_{d} \times \mathbb{S}^{D-d}.
\end{equation}
This spacetime geometry could be interpreted as the near horizon geometry of $(d-2)$ branes in such  superstring/M-theory inspired models \cite{Adil1}. An examination on the sphere compactification shows that such higher dimensional models should,  at least, involve a $(D-d)$ strength gauge field $\mathcal{F}_{D-d}$ contributing with the term $\int \mathrm{d}^d x \, \mathcal{F}_{D-d}^{2}$ in the associated lower dimensional black hole action. The presence of such a term is supported by a $(D-d-1)$ gauge form coupled to a $(D-d-2)$-brane supposed to live in such higher dimensional inspired models. After a close inspection, it has been remarked that  one has two possible distinct  brane objects  in the proposed models. They could be classified as
\begin{itemize}
\item $ \,(d-2)$-branes associated with the $AdS_d$ geometry of the  black hole,
\item $ \,(D-d-2)$-branes corresponding to the $\mathbb{S}^{D-d}$ sphere compactification.
\end{itemize}
In the study of such higher dimensional inspired models, two cases could arise
\begin{enumerate}
\item $ D-d-2 = d-2$ leading to $D=2d$,
\item $ D-d-2 \neq d-2$ giving $D \neq 2d$.
\end{enumerate}
It is worth noting that particular cases appear in known theories associated with $D=10$ and $D=11$. For instance, the first case arises in type IIB superstring in the presence of $D3$-branes \cite{I25}. However, the second situation  takes place in M-theory involving $M2$ and $M5$-branes \cite{I26,I27}. Roughly,  the two previous conditions can be put in one relation given by
\begin{equation}
D=2d+k,
\label{dim}
\end{equation}
where now $k$ is an arbitrary integer which will be used to specify the internal space of $(D-d)$ dimensions. In this notation, the $d$-dimensional AdS black holes are obtained by the compactification of the $D$-dimensional theory on the real spheres  $\mathbb{S}^{d+k}$. 
The resulting models will be classified by a triplet $(D,d,k)$ subject to the relation \eqref{dim}. Using this notation, the electric-magnetic duality is assured by the transformation
\begin{equation}
(D,d,k) \longleftrightarrow (D,d+k,-k).
\end{equation}
In terms of the $AdS$ geometry, this duality can be rewritten as
\begin{equation}
AdS_{d} \times \mathbb{S}^{d+k} \longleftrightarrow AdS_{d+k} \times \mathbb{S}^{d}.
\end{equation}

Well known theories correspond to   special choices of the triplet $(D,d,k)$. 
For instance, the space-time of type IIB superstring theory associated with the triplet $(10,5,0)$ is given by $ \left( AdS_{5} \right)_{L} \times \left( S^{5}\right)_{L}$ linked  to  $D3$-brane physics. For M-theory, one has the triplet $(11,4,3)$ described by $ \left( AdS_{4} \right)_{L/2} \times \left( S^{7}\right)_{L}$ based on $M2$-branes being 
dual to the triplet $(11,7,-3)$ associated with the $ \left( AdS_{7} \right)_{2L} \times \left( S^{4}\right)_{L}$ geometry relying on $M5$-branes.

The near horizon geometry of the $(d-2)$-branes spacetime manifold   becomes now  
\begin{equation}
AdS_{d} \times \mathbb{S}^{d+k},
\end{equation}
with the associated line element given by 
\begin{equation}
ds^2=-f(r)dt^2+\frac{1}{f(r)}dr^2  +r^2h_{ij}dx^idx^j+
L^2d\Omega^2_{d+k}.
\end{equation}
It is noted that  $h_{ij} \, dx^idx^j$ is the line element of a $(d-2)$-dimensional
Einstein space $(\Xi_{d-2})$ \cite{I28}. 
The quantity $d\Omega^2_{d+k}$  is  the metric of the
$\mathbb{S}^{d+k}$ real sphere with radius $L$. 

In a  AdS/CFT context,   this radius is linked to the  brane number \cite{I29}. 
The function $f(r)$  depends on physical parameters including the possible existence of 
 non trivial backgrounds, i.e. quintessence. 
Such a situation will be dealt with  in the present study. 

%
%

%
%
%
%

Using the  well established procedures (see \cite{I23,I24,notrea},\cite{Liu:2017baz,pengzhao,Wu:2020tmz}), 
we can check that a line element metric with the 
 $f(r)$ function of the form 
\begin{equation}
f(r)= 1-\frac{m}{r^{d-3}}-\sum_{n} \left( \frac{r_{n}}{r} \right)^{(d-1)\omega_{n}+(d-3)},
\label{e002}
\end{equation}
where $\omega_n$ are free parameters and $r_{n}$ are dimensional normalization constants,   is 
a solution of the Einstein Equations for a suitable  energy-momentum tensor.

Some well known situations are particular cases of (\ref{e002}). For  instance,  one can consider the case 
\begin{equation}
f(r)=1-\frac{m}{r^{d-3}}-\frac{c}{r^{(d-1)\omega_{q}+(d-3)}},
\end{equation}
where $m$ is an integration constant and $c$ represents the DE contributions \cite{I30,s26}. 
The associated quintessence energy density  $\rho_q$  is  given by
\begin{equation}
\rho_q=-\frac{c \, \omega_q(d-1)(d-2)}{4r^{(d-1)(\omega_q+1)}}.
\end{equation}
For $\omega_0=-1$, we obtain
\begin{equation}
f(r)= 1-\frac{m}{r^{d-3}}+ \left( \frac{r}{L} \right)^2,
\end{equation}
producing an AdS-Schwarzschild black hole solution in $d$-dimensions. However, the spherical Schwarzschild solution is obtained by taking the large limit of AdS radius $(L^2 \rightarrow \infty)$.  Another case associated with the   $d$-dimensional Reissner-Nordstrom black hole   is obtained  by  taking  $ \omega_1=\frac{d-3}{d-1}$
\begin{equation}
f(r)= 1-\frac{m}{r^{d-3}}+ \left( \frac{r}{L} \right)^2+\frac{Q^2}{r^{2(d-3)}}
\end{equation}
where $Q$ denotes the associated charge. A close inspection  shows that  a  $d$-dimensional AdS-Schwarzschild black hole, in the presence of  quintessence,  can be described  by the following metric function
\bea
f(r)&=&1-\frac{m}{r^{d-3}}+\frac{r^{2}}{L^{2}}-\frac{c}{r^{(d-1) \omega_{q} + d-3 }},
\label{f}
\eea
where $c$ is a positive normalization factor associated with DE intensity given by $r_q^{(d-1)\omega_q+(d-3)}$.

This  last case will treated in full detail in the next sections.

\subsection{ Thermodynamics of  $d$-dimensional AdS-Schwarzschild black hole  in the presence of  quintessence }

A $d$-dimensional AdS-Schwarzschild black hole, in the presence of  quintessence with parameter $\omega_q$,
  can be described  by the 
 metric function given in Eq.\eqref{f}.
The event horizon $r_{h}$ in this case, is determined by setting Eq.(\eqref{f}) equal to zero $(f(r)=0)$. This gives as solution for 
the coefficient $m$ the value
\begin{equation}
m=r_h^{d-3}+\frac{r_h^{d-1}}{L^{2}}-c \, r_h^{\omega_{q}(1-d)}.
\label{m}
\end{equation}
It turns out that the general form of the black hole mass reads as
\begin{equation}
M_{d}=\frac{(d-2) \, \varpi_{d-2} }{16 \pi \, G_{d}} \, m,
\end{equation}
where the factor $\varpi_{d-2}$ is given by
\begin{equation}
\varpi_{d-2}=\frac{2 (\pi)^{\frac{d-1}{2}}}{\left( \frac{d-3}{2} \right)! },
\label{omegad2}
\end{equation}
identified with the volume of the Einstein space $\Xi_{d-2}$ \cite{s28}. It is recalled that $G_{d}$ is the gravitational constant \cite{s28,s29}.
Using Eq.\eqref{m}, we get the mass expression
\begin{equation}
M_{d}=\frac{(d-2) \, \varpi_{d-2}}{16 \pi \, G_{d}} \left( r_h^{d-3}+\frac{r_h^{d-1}}{L^{2}}-c \, r_h^{\omega_{q}(1-d)} \right).
\label{MG}
\end{equation}
For $d$-dimensional AdS black holes \cite{s210}, the general formula of Bekenstein-Hawking entropy takes the following form
\begin{equation}
S_{d}=\frac{\varpi_{d-2} \,r_h^{d-2}}{4 G_{d} }.
\label{entropy}
\end{equation}
In terms of such  an entropy, $r_h$ is given by
\begin{equation}
r_h= \left( \frac{4 G_{d}}{\varpi_{d-2}} \right)^{\frac{1}{d-2}} S^{\frac{1}{d-2}}.
\label{Rg}
\end{equation}
The gravitational constant $G_d$ in  such a $d$-dimensional AdS black hole should 
be related to the one corresponding to $(2d+k)$-dimensional inspired models. It is 
noted that $d$-dimensional AdS black hole theory can be obtained by the compactification of 
the $(2d+k)$-dimensional theory on the $\mathbb{S}^{d+k}$ sphere of radius $L$ \cite{I13}. 
From such  dimensional spherical reductions, we get
\begin{equation}
G_{d}=\frac{G_{2d + k}}{\text{Vol}\left( \mathbb{S}^{d+k} \right)}=\frac{G_{2d + k}}{ \omega_{d+k} \, L^{d+k}},
\label{cteGg}
\end{equation}
where $\omega_{d+k} $ is given by
\begin{equation}
\omega_{d+k} = \frac{2 \pi^{(d+k+1)/2}}{\Gamma(\frac{d+k+1}{2})}.
\label{omegaDd}
\end{equation}

%

After a close inspection, we show that the radius $L$ verifies
\begin{equation}
L^{2(d-1)+k}=2^{-\left( \frac{d\left(4-d\right)+3}{2}  \right)} \, \pi^{7\left( k+2(d-5) \right)-4} \, N^{\frac{d-1}{2}} \ell_{p}^{ \, {2(d-1)+k}}
\label{2.11}
\end{equation}
where  $\ell_p$ and $N$ are  the Planck length   and the   brane number,  respectively.

By putting Eqs.\eqref{omegad2},\eqref{Rg} and \eqref{cteGg} 
in Eq.\eqref{MG}, the general mass form reads as
\begin{equation}
\begin{aligned}
M^{(k)}_d(S,N, c)&  =  \frac{\left(d-2 \right)}{8} \,  \frac{\pi^{\left(\frac{d-3}{2} \right)}}{\left( \frac{d-3}{2} \right)!} \frac{\ell_{p}^{d+k-2} \, \omega_{d+k} \, B^{\frac{d+k-2}{2}}}{G_{2d+k}}  \\
& \quad \times  \left\lbrace \ell_{p}^{2} \, B \left[ A^{\frac{d-3}{d-2}} -c A^{\frac{-\omega_{q}(d-1)}{d-2}} \right] + A^{\frac{d-1}{d-2}} \right\rbrace,
\end{aligned}
\label{Mgf}
\end{equation}
where  $A$ and $ B$ are given by
\bea
B(N)&=&2^{\frac{d(d-4)-3}{k+2(d-1)}} \, \pi^{\frac{2\left( 7 \left(k+2(d-5)\right) -4 \right)}{k+2(d-1)}} \, N^{\frac{d-1}{k+2(d-1)}},\\
 A(S,N)&=&\frac{\left( \frac{d-3}{2} \right)!}{\pi^{\frac{d-1}{2}}} \cdot \frac{\left(2S \right) G_{2d+k} \,}{ \omega _{d+k} \,  \ell_p^{d+k} \, B^{\frac{d+k}{2}}}.
\label{B}
\eea

Exploiting the  first law of black hole thermodynamics
\begin{equation}
dM =T dS + \mu dN^{\frac{d-1}{2}}
\end{equation}
and using Eq.\eqref{Mgf}, we find the associated thermodynamic temperature $T$
\begin{equation}
\begin{aligned}
T^{(k)}_d(S,N, c)& = \frac{\partial M^{(k)}_d}{\partial S} \bigg\vert_{N} =  \frac{1}{8 S} \,  \frac{\pi^{\left(\frac{d-3}{2} \right)}}{\left( \frac{d-3}{2} \right)!} \, \frac{\ell_{p}^{d+k-2} \, \omega_{d+k} \, B^{\frac{d+k-2}{2}}}{G_{2d+k}}  \\
& \times  \left\lbrace \ell_{p}^{2} \, B \left[ \left(d-3 \right) A^{\frac{d-3}{d-2}} +c \, \omega_{q} \left(d-1\right) A^{\frac{-\omega_{q}(d-1)}{d-2}} \right] + \left( d-1 \right) A^{\frac{d-1}{d-2}} \right\rbrace
\end{aligned}
\label{Tgf}
\end{equation}
and the chemical potential $ \mu $,   being conjugate variable associated with  the number 
of colors $N^{(d-1)/2}$, as
\begin{equation}
\mu^{(k)}_d(S,N,c)  = \frac{\partial M^{(k)}_d}{\partial N^{\frac{d-1}{2}}} \bigg\vert _{S}=\frac{2}{(d-1) \cdot N^{\frac{d-3}{2}}} \left( \frac{\partial M^{(k)}_d}{\partial N} \right) \bigg\vert_{S}.
\end{equation}
A simple calculation gives
\bea
 \mu^{(k)}_d(S, N, c)   =&&
  \frac{1}{8N^{\frac{d-1}{2} }} \, \frac{1}{k+2(d-1)} \, \frac{\pi^{\left(\frac{d-3}{2} \right)}}{\left( \frac{d-3}{2} \right)!} \, \frac{\ell_{p}^{d+k-2} \, \omega_{d+k} \, B^{\frac{d+k-2}{2}}}{G_{2d+k}}   \\ \nonumber
 \times  && \left\lbrace \ell_{p}^{2}  (d+k) B
\left[ 
A^{\frac{d-3}{d-2}} \right. \right.- \left. \left. c   ((d-2)+(d-1) w_q)
A^{\frac{-\omega_{q}(d-1)}{d-2}} \right] 
+ 
(3d +k-4) A^{\frac{d-1}{d-2}} 
\right\rbrace.
\label{mugf} \nonumber
\eea
%
The Gibbs (or Helmholtz)  free energy can be also computed by applying the following relation
\begin{equation}
G^{(k)}_d(S,N,c)=M^{(k)}_d(S,N,c)-T^{(k)}_d(S,N,c) \cdot S.
\end{equation}
Using Eqs \eqref{Mgf} and \eqref{Tgf}, we get
\begin{equation}
\begin{aligned}
G^{(k)} _d(S,N, c)& =  \frac{\pi^{\left(\frac{d-3}{2} \right)}}{\left( \frac{d-3}{2} \right)!} \, \frac{\ell_{p}^{d+k-2} \, \omega_{d+k} \, B^{\frac{d+k-2}{2}}}{8 \,
 G_{2d+k}}  \\
& \times  \left\lbrace \ell_{p}^{2} \, B \left[ A^{\frac{d-3}{d-2}} -c \, \left[ \left( \omega_{q}+1 \right) \left( d-1 \right)-1 \right] A^{\frac{-\omega_{q}(d-1)}{d-2}} \right] - A^{\frac{d-1}{d-2}} \right\rbrace.
\end{aligned}
\label{Ggf}
\end{equation}
%
%
%

\section{The baseline case: Hawking-Page phase transitions without dark energy}

We consider first the Hawking-Page phase transition in the absence of DE (the case 
 $c = 0$ in the previous section) 
in different models. These models  will  be denoted by the triplet $(D,d,k)$. 
Fixing the 
 space-time dimension $D$, the thermodynamical quantities will be indicated by two numbers $d$ and $k$. The thermodynamical  quantities will be referred to as $X_d^{(k)}$. In absence of DE, 
 $X_d^{(k)}$ depend only  on the brane number $N$ and  the entropy $S$. 


Here, we  deal with  a model associated with  the compactification of 11-dimensional 
 M-theory in the presence of $N$ coincident $M5$-branes, indexed by  the triplet
\begin{equation}
(D,d,k)=(11,7,-3).
\end{equation}
This  corresponds to the  $AdS_{7}\times \mathbb{S}^{4}$ space-time, where the gravitational constant in eleven dimensions is $G_{11}=2^4 \pi^7 \ell_{p}^{9}$. 
In this case, the internal space is the four-dimensional real sphere $\mathbb{S}^4$ with the 
 volume factor
\begin{equation}
\omega_{4}=\frac{8 \pi^{2}}{3},
\label{omegaAds7}
\end{equation}
obtained from Eq.\eqref{omegaDd}. This compactification produces a seven-dimensional AdS black hole. Using  Eq.\eqref{omegaAds7}, we compute the  corresponding thermodynamical quantities $X_d^{(k)}$ denoted by the doublet $(7,-3)$. They are listed as follows
\begin{itemize}
\item mass
\begin{equation}
M_{7}^{(-3)}(S,N)=\frac{5\left[ 8 \times 3^{3/5} \, \pi^{2/5} \,  S^{\frac{4}{5}} \, N^{\frac{4}{15}} + 3 \times  2^{3/5} \, S^{\frac{6}{5}} N^{\frac{-14}{15}} \right]}{16 \times   6^{4/5} \, 3^{1/5} \,  \pi^{23/15} \, \ell_{p}},
\label{M70}
\end{equation}

\item Hawking temperature
\begin{equation}
T_{7}^{(-3)}(S,N)=\frac{  16 \times 3^{3/5} \, \pi^{2/5} \,  S^{-\frac{1}{5}} \, N^{\frac{4}{15}} + 9 \times  2^{3/5} \, S^{\frac{1}{5}} N^{\frac{-14}{15}} }{8 \times   6^{4/5}  \,  \pi^{23/15} \, \ell_{p}},
\label{T70}
\end{equation}

\item chemical potential
\begin{equation}
\mu_{7}^{(-3)}(S,N)=\frac{ 16 \times 3^{3/5} \, \pi^{2/5} \,  S^{\frac{4}{5}} \, N^{-\frac{11}{15}} -21 \times  2^{3/5} \, S^{\frac{6}{5}} N^{\frac{-29}{15}} }{24 \times   6^{4/5} \,  \pi^{23/15} \, \ell_{p}},
\label{mu70}
\end{equation}
\item Gibbs free energy
\begin{equation}
G_{7}^{(-3)}(S,N)=\frac{ 8 \times 3^{3/5} \, \pi^{2/5} \,  S^{\frac{4}{5}} \, N^{\frac{4}{15}} - 3 \times  2^{3/5} \, S^{\frac{6}{5}} N^{\frac{-14}{15}} }{16 \times   6^{4/5}  \,  \pi^{23/15} \, \ell_{p}}.
\label{G70}
\end{equation}
\end{itemize}

\begin{figure} 
\centering
\includegraphics[scale=0.55]{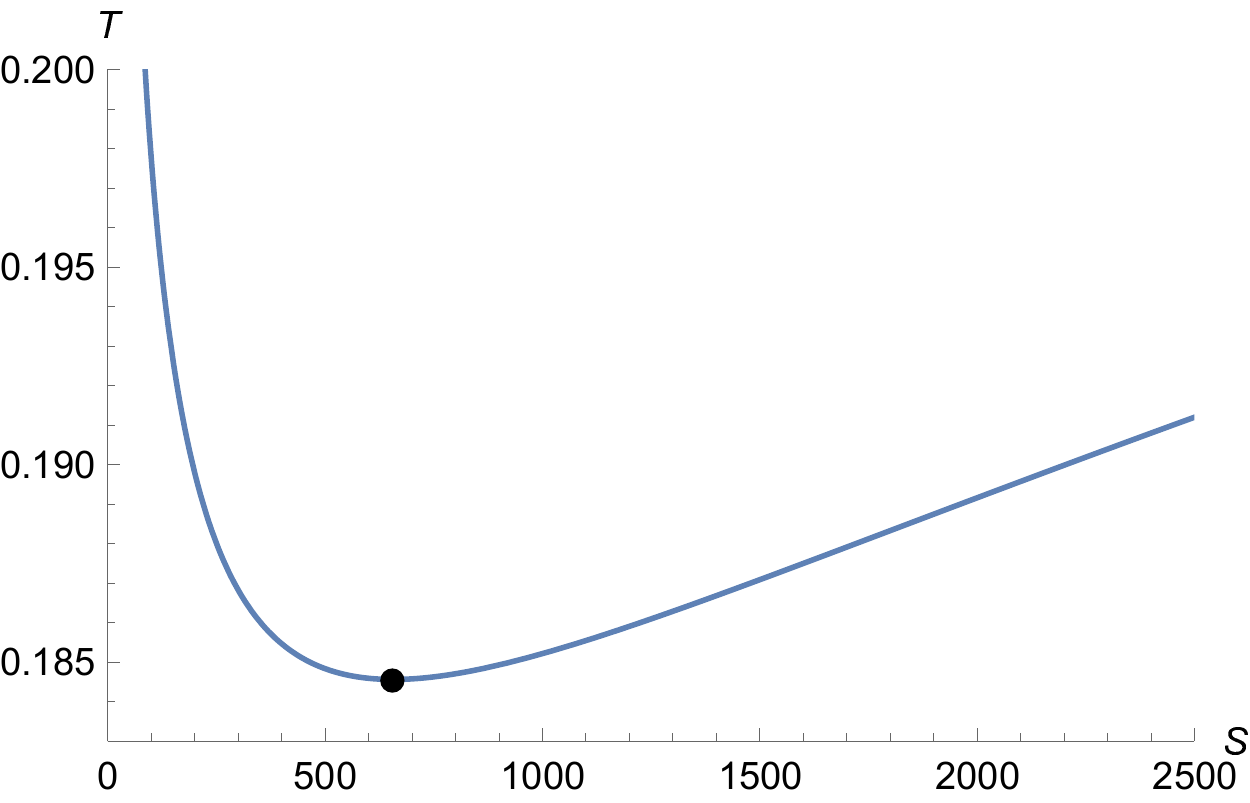}
\caption{Temperature as function of the entropy for fixed $N=3$ and $\ell_{p}=1$.
The temperature reaches its minimum at $S_{7}^{min} \simeq 657$.}
\label{T7}
\end{figure}

In order to investigate the behaviour of the system in presence of the 
Hawking-Page  transitions, we first consider  the dependence of the
 Hawking temperature with respect to the entropy. This is illustrated in figure \ref{T7}.
For a generic number $N$ of $M5$-branes, 
the Hawking temperature has a minimum for
\begin{equation}
S_7^{min}=\frac{2^{17/2}}{3^{7/2}} \pi N^{3},
\end{equation}
 corresponding  to the 
minimal temperature
\begin{equation}
T_7^{min}=\frac{3^{1/2}}{2^{1/2} \, \pi ^{4/3} N^{1/3} \ell_{p}}.
\end{equation}
Below such a   minimal temperature, no black hole solution can exist. However, it follows from figure \ref{T7} that above this temperature one can distinguish two branches. One corresponding to a small entropy is associated with a thermodynamically unstable black hole. The second branch, with a large entropy, describes a thermodynamically 
stable black hole.

\begin{figure}[t]
\begin{tabular}{ll}
\includegraphics[scale=0.4]{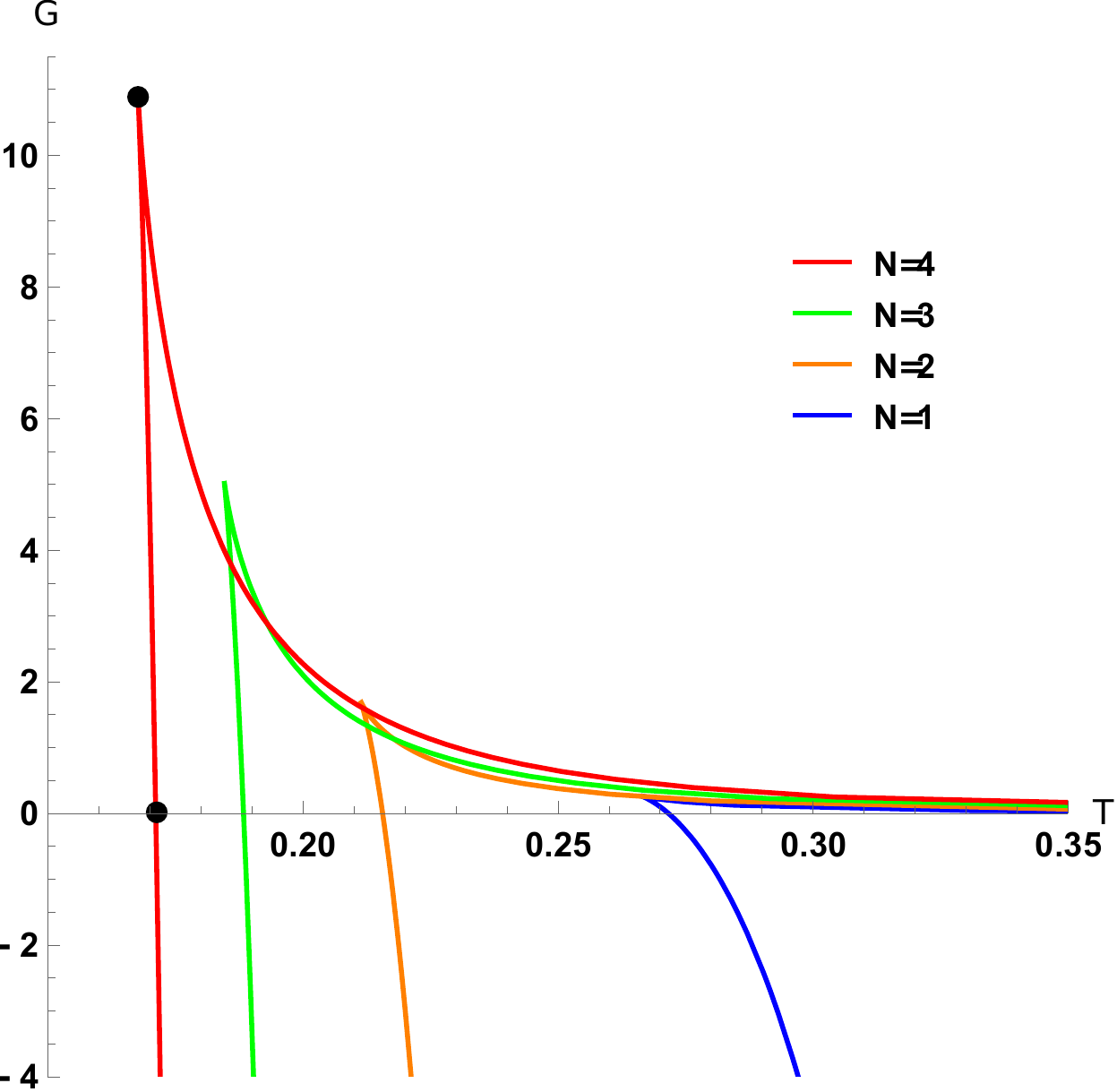} &\includegraphics[scale=0.55]{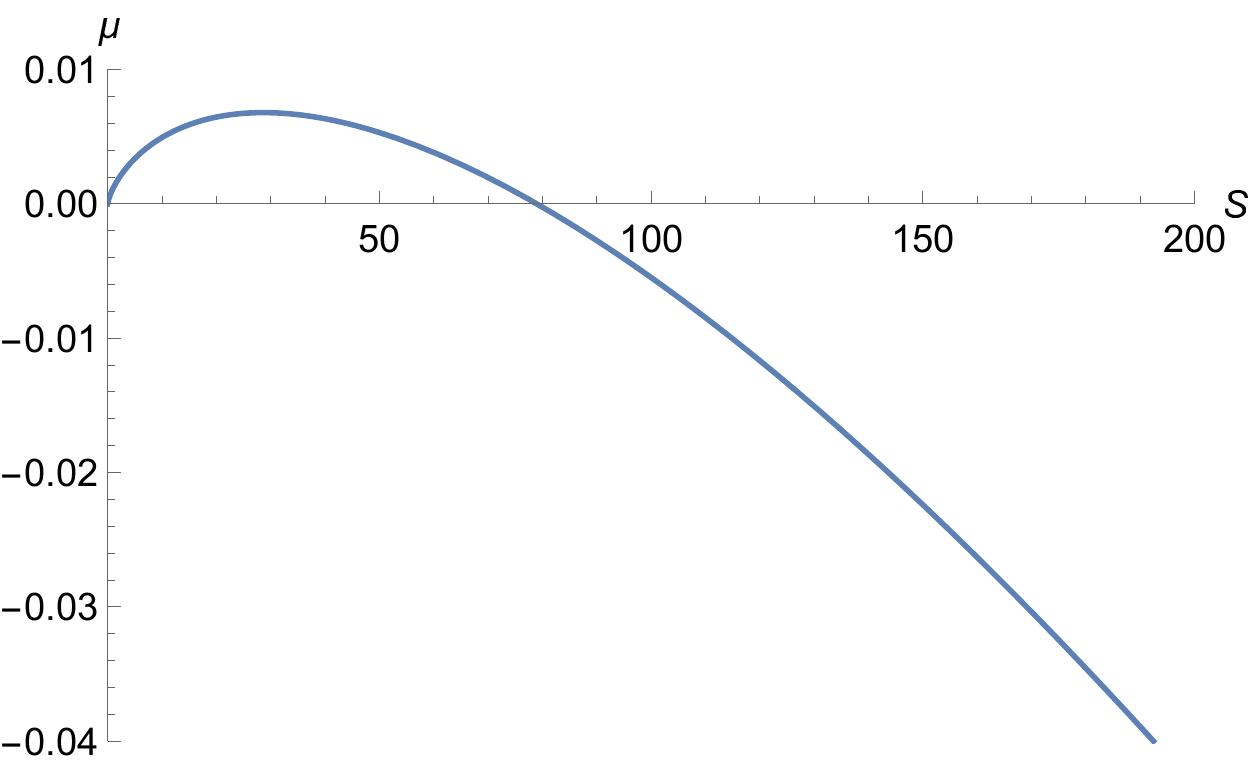}\\
\includegraphics[scale=0.55]{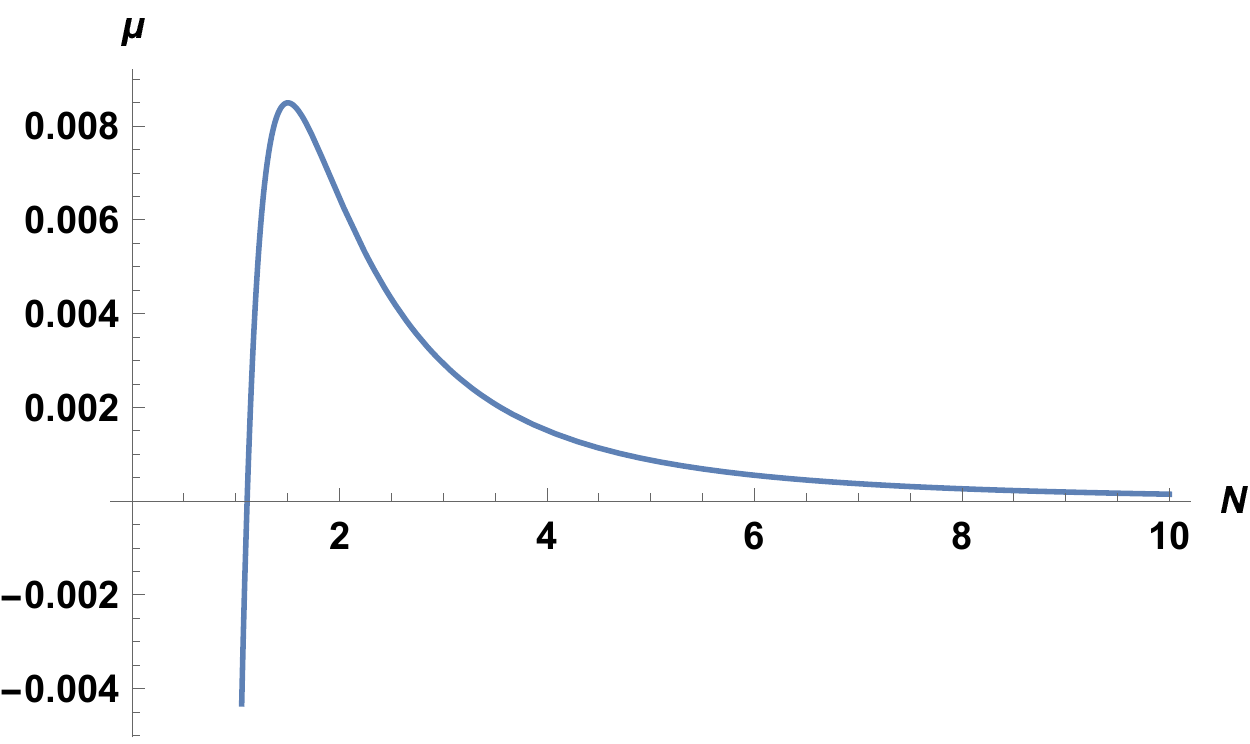} & \includegraphics[scale=0.4]{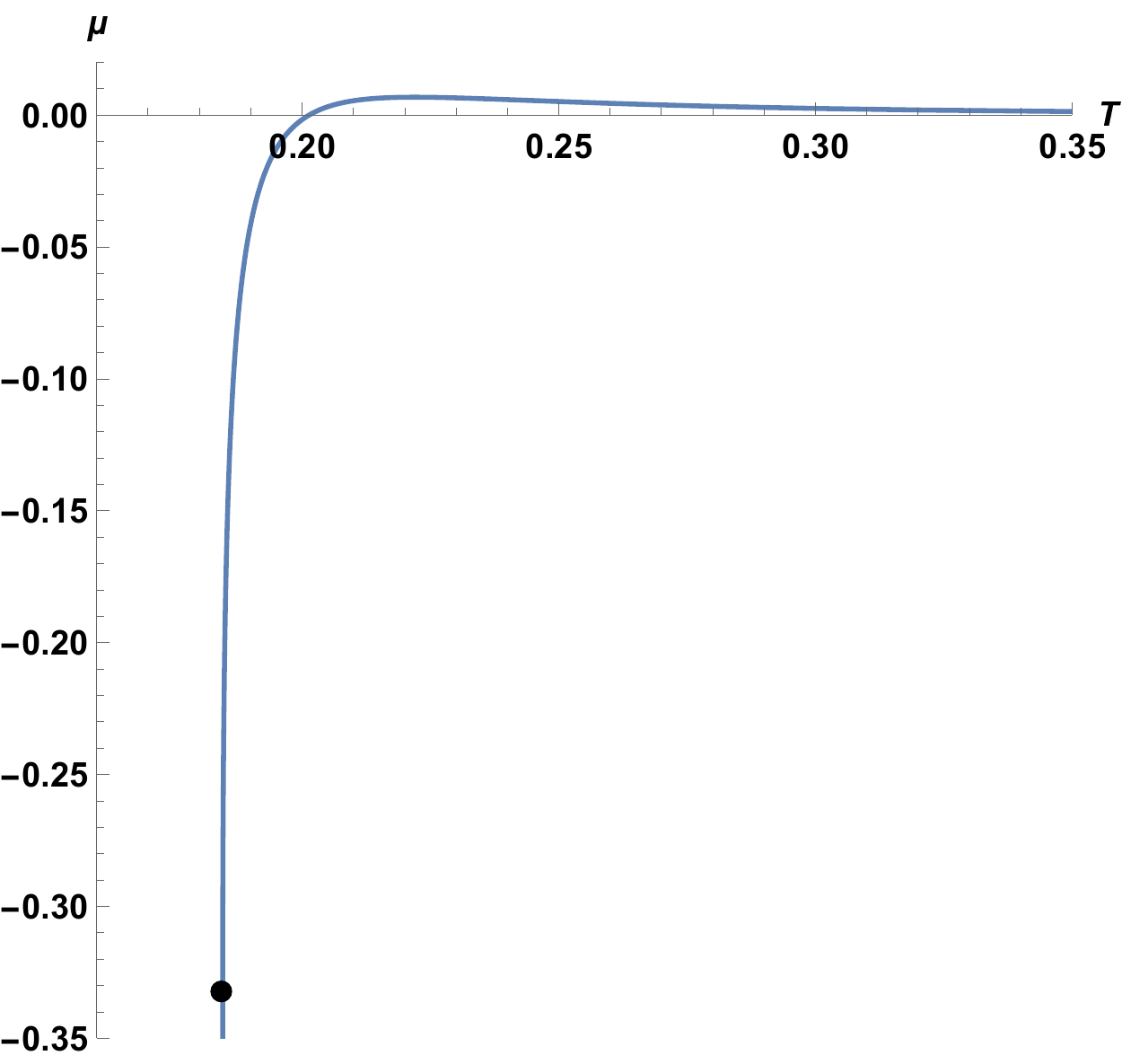}
\end{tabular}
\caption{
(Top Left) The Gibbs free energy as function of temperature for different values of $N$ 
(fixed $\ell_{p}=1$).  The Hawking-Page phase transition temperature $T^{HP}$ is 
located at the lower dot (over the $x$-axis), $T^{min}$  is located at the upper dot. 
(Top Right)
The chemical potential as a function of entropy for $N = 3, \ell_{p}=1$. 
It changes sign   at $S_7^{\mu=0} \simeq 79$.
(Bottom Left) 
The chemical potential as a function of the number of $M5$-branes $N$. Here we take $\ell_{p} = 1$ and $S = 4$. 
The maximum occurs at  $N_7^{max} \simeq 1,5$ for $S_7^{max}=4$.
(Bottom Right)
The chemical potential as a function of temperature $T$. We take $\ell_{p} = 1$ and $N = 3$.}
\label{mu7T}
\end{figure}

The sign of the Gibbs free energy changes at the Hawking-Page temperature $T_7^{HP}$ 
associated with $S_7^{HP}$. Taking into account the Gibbs free energy given in Eq.\eqref{G70}, this phase transition occurs at ($G(T^{HP},N)=0$, $S^{HP}=S_{G=0}$)
\begin{equation}
S_7^{HP}=\frac{2^{6}}{3} \pi N^{3},
\end{equation}
corresponding to  the Hawking-Page temperature
\begin{equation}
T_7^{HP}=\frac{5}{4 \, \pi ^{4/3} N^{1/3} \ell_{p}},
\end{equation}
which is always greater than $T_7^{min}$.
The dependence of the brane number $N$ on the Hawking-Page phase transition of the 
$AdS$ black hole is represented in figure \ref{mu7T}. In particular,  we plot the Gibbs free 
energy as a function of the Hawking temperature $T_7^{(-3)}$ for different values of $N$. 
In this plot,  it becomes apparent that below $T_7^{min}$ (the higher black dot)  no 
thermodynamically stable black hole can survive.
The Gibbs free energy reaches a maximum at $T_7^{min}$. For lower Gibbs free energy values, 
one finds two branches. The upper branch describes a small (unstable) black hole with a negative 
specific heat. The lower branch corresponds to (large) stable black hole solutions with positive specific 
heat values. The crossing point in this branch for which $G=0$  corresponds to 
$T_7^{HP}$. At this point,  
the (first order) Hawking-Page phase transition  occur 
 between large (stable) black holes and a thermal radiation state \cite{I7}.
In figure \ref{mu7T}, we plot the chemical potential as a function of the entropy $S$ for a fixed number 
of $M5$-branes.
The chemical potential is positive for smaller entropy $S$, and negative for large
values.  The sign change in the chemical potential occurs at the  entropy value
\begin{equation}
S_7^{\mu=0}=\frac{2^{17/2}}{3 \cdot 7^{5/2}} \, \pi N^3
\label{s7mu0}
\end{equation}
 which corresponds the temperature
\begin{equation}
T^{\mu=0}_7=\frac{5}{\sqrt{14} \pi^{4/3} \ell_p N^{1/3}}.
\label{t7mu0}
\end{equation}

We have  $S_7^{\mu=0} < S_7^{min} < S_7^{HP}$. The entropy where the chemical potential changes the sign is less than the one at which no black hole can exist. 
 The chemical potential dependence as a function of $N$ with a fixed entropy $S$
is showed in  \ref{mu7T}.
It follows from figure \ref{mu7T} that the maximum of the chemical potential corresponds to the entropy 
$$ S_7^{max} = \frac{2^{17/2}}{3 \cdot 7^{5/2}} \, \left( \frac{41}{59} \right)^{5/2} \, \pi \left( N_7^{max} \right)^3.$$
In figure \ref{mu7T}, we plot the chemical potential as a function 
of the temperature $T$ for a fixed number of $M5$-branes $N$.
It is noted that the dot, appearing in figure \ref{mu7T}, indicates the minimum of the temperature $T^{min}_7$. However, bellow this point one can find the Hawking-Page temperature $T_7^{HP}$, separating the stable lower branch (very low values of the chemical potential)  and the upper unstable branch where $T_7^{min}$ resides. From Eq.\eqref{s7mu0}, the chemical potential is positive for the brane number condition $N^3 > \frac{3 \cdot 7^{5/2} }{2^{17/2}} \, \frac{S}{\pi}$. This limit can be saturated for the temperature
\begin{equation}
T^{\mu=0}_7 \simeq 1.1\, T_7^{HP}.
\end{equation}
Similarly as in \cite{X0}, we see that $T^{\mu=0}_7 > T_7^{HP}$ which means that the black hole is preferred over pure $AdS_7$ backgrounds.

%
%


\subsection{The Darkless case: further models}

Similar results are obtained for other cases.  In particular, the results for $(D,d,k)=(11,4,3)$ and $(D,d,k)=(10,5,0)$ will be  presented in appendices A and B, respectively. 
A summary of different quantities associated to each model is shown in 
 table \ref{Tableau}.

\begin{table}[t]
\begin{center}
\begin{tabular}{|l||l|l|l|l|l|l|l|l|l|}
\hline
$AdS_{d}\times \mathbb{S}^{d+k}$ & $S_d^{min}$ & $T_d^{min}$ & $S_d^{HP}$ & $T_d^{HP}$ & $S_d^{\mu=0}$ & $S_d^{max}$ & $N_{max}$ & $G_{min}$\\
\hline
\hline
$AdS_{4}\times \mathbb{S}^{7}$ & $0.07$ & $0.18$ & $0.2$ & $0.2$ & $0.1$ & $0.05$ & $55$ & $0.004$\\
\hline
$AdS_{5}\times \mathbb{S}^{5}$ & $10$ & $0.5$ & $28$ & $0.6$ & $8.6$ & $3.5$ & $3.2$ & $0.7$\\
\hline
$AdS_{7}\times \mathbb{S}^{4}$ & $657$ & $0.19$ & $1800$ & $0.19$ & $79$ & $31.7$ & $1.5$ & $5$\\
\hline
\end{tabular}
\caption{Summary of results of different cases. ( $N=3$  and $\ell_{p}=1$)}
\label{Tableau}
\end{center}
\end{table}

From this table, one can observe some systematic  features among the models.
First, we notice an increasing behavior for the entropy as increases  the dimension $d$, the dimension in which the $AdS$ black holes live. This can be seen clearly in the transition 
points listed in table \ref{Tableau}, figures \ref{T7}, \ref{T4} and \ref{T5}. 
This can simply 
be understood from the extensive character of  
 the Bekenstein Hawking entropy.
%
%
%
%
%
%
However, although increasing with the dimension, from the table \ref{Tableau}, 
we remark   the universal character of the normalized ratio
$\frac{S_d^{min}}{S_d^{HP}}$, in fact in all the cases, at different space-time dimensions,
\begin{equation}
\frac{S_d^{min}}{S_d^{HP}} \simeq \frac{1}{3}.
\end{equation}
Independently of the dimension $d$, the entropy where the large black holes transit to thermal $AdS$ is approximately three times the entropy for which there is no black hole solution. This relation seems to be a universal one.

Second, one can notice an obvious decrease in the number of branes 
($N_{max}$, the value for which the chemical potential is maximum for fixed entropy)  as 
the dimension $d$ of the $AdS$ space increases: the four-dimensional $AdS$ space 
"needs"  a larger  number of $M2$-branes in comparison 
 to the seven-dimensional $AdS$ space with $M5$-branes in the context of the M-theory compactification.

Third, the Hawking temperature of five-dimensional $AdS$ black hole in type IIB superstring is bigger than the one associated with other AdS black holes appearing in M-theory. Taking a particular case of the ratio $T_5^{min}/T_7^{min}$ which is equal to
\begin{equation}
\frac{2^{7/8} \pi^{5/6} N^{1/12}}{3^{1/2}},
\end{equation}
we find $T_5^{min} > T_7^{min}$. 
Since, below $T_d^{min}$ no black hole can exist, we see that this 
phase is larger for $AdS_5$ followed by $AdS_7$ and $AdS_4$. 
However, the radiation phase of $AdS_5$ is larger than the radiation phase of $AdS_4$. 
The smaller radiation phase correspond to $AdS_7$ case. 
Taking the ratio $T_4^{HP}/T_7^{HP}$ for the particular case of $N=1$, we find the opposite behavior.

%

\section{Hawking-Page phase transitions in presence of dark energy}

In this section, we investigate the effect of DE surrounding  AdS black holes. 
The DE state parameters are the free parameter $c$ (see Eq.\ref{f})  and 
 $\omega_q=\frac{p}{\rho}$ defined in terms of a ratio of the  pressure and the  density of DE in the universe \cite{I32,I33,I34}. Many forms of such energy have been proposed depending on particular values of $\omega_q$. Among others, quintessence associated with a dynamic field has been extensively investigated including string theory and brane physics \cite{I35,I36}.
 Here, such DE contributions will be approached in the context of $AdS$ black holes in M-theory/IIB superstring inspired models.
 
We first  compute the AdS black hole thermodynamical quantities like the mass, the temperature, the Gibbs free energy and the chemical potential. This will be made using the general relations obtained in section \ref{general}. 
Then, we discuss the effect of the quintessence on the associated phase transitions.
The previous thermodynamical quantities $X_d^{(k)}(N,S)$ will be replaced by $X_d^{DE \, (k)}(N,S,c)$ 
where $c$ represents the DE contributions.  The parameter $w_q$ will be fixed by dimensional considerations for any of the models.

We discuss in detail a model denoted  by the triplet $(D,d,k)=(11,7,-3)$, i.e.  a compactification of M-theory on the sphere $\mathbb{S}^4$ in the presence of $M5$-branes: $AdS_ 7\times \mathbb{S}^{4}$.
Further results for $AdS_ 4\times \mathbb{S}^{7}$ and $AdS_ 5\times \mathbb{S}^{5}$ spacetimes will be  presented in the appendices C and D,  respectively.

Using Eq.\eqref{omegaAds7}, we get the following expressions for thermodynamic variables of interest
\begin{itemize}
\item mass
\begin{equation}
M_{7}^{(-3) \, DE}(S,N,c)=M_{7}^{(-3)}(S,N) -\alpha^{DE}_7(S,N,c),
\label{MDE70}
\end{equation}

\item temperature
\begin{equation}
T_{7}^{(-3) \, DE}(S,N,c)=T_{7}^{(-3)}(S,N) +\frac{6\omega_q}{5} \cdot \frac{\alpha^{DE}_7(S,N,c)}{S},
\label{TDE70}
\end{equation}

\item chemical potential
\begin{equation}
\mu_{7}^{(-3) \, DE}(S,N,c)=\mu_{7}^{(-3)}(S,N) - \frac{4 \left( 6\omega_q + 5 \right)}{45} \cdot \frac{\alpha^{DE}_7(S,N,c)}{N^{2}},
\label{muDE70}
\end{equation}
\item Gibbs free energy
\begin{equation}
G_{7}^{(-3) \, DE}(S,N,c)=G_{7}^{(-3)}(S,N) - \left( \frac{6\omega_q+5}{5} \right)\cdot \alpha^{DE}_7(S,N,c).
\label{GDE70}
\end{equation}
\end{itemize}
For all these quantities, $\alpha^{DE}_7(S,N,c)$ denotes the DE contribution which is given by
\begin{equation}
\alpha^{DE}_7(S,N,c)=\frac{5 \, c \, 2^{\frac{6\omega_q-5}{5}} N^{\frac{4(6 \omega_q+5)}{15}}}{ 3^{\frac{6\omega_q+5}{5}} \pi^{\frac{12 \omega_q +25}{5}} \, S^{\frac{6 \omega_q}{5}}  \ell_p^{6\omega_q+5}}.
\label{alpha7}
\end{equation}
It is noted that  $c$ is a free parameter (in a certain range), the quantity $w_q$ is restricted by dimensional analysis.
In associated  black hole physics,  the thermodynamical quantities  should be proportional to the Planck mass $m_{p}$.  Indeed, an examination shows that   such  quantities are all proportional to $\ell_p^{-1}$. In the $AdS_7 \times \mathbb{S}^4$  for  instance,   we have 
\begin{equation}
[M_7^{(-3) \, DE }]=[M_7^{(-3) }]=[ \alpha_7^{(-3)}].
\end{equation}

Since $M_7^{(-3) \, DE }$ should be proportional to $\ell_p^{-1}$, $\alpha_7^{(-3)}$ should  have the same physical dimension. Using the expression of $\alpha_7^{(-3)}$ given in Eq.\eqref{MDE70}, we find that this quantity is proportional to $\ell_p^{-(6\omega_q+5)}$. A simple calculation  gives 
\begin{equation}
 \omega_q=-2/3.
\end{equation}
Other values $\omega_q$ will be obtained for $S^7$ and $S^5$ compactifications (see  appendices C and D,  respectively).
In this way,  (\ref{alpha7}) reduces to 
\begin{equation}
\alpha_7^{DE} \left(S,N,c\right) = \frac{5 \, c \, 2^{-9/5}
N^{4/15}}{3^{1/5} \pi^{17/15} \, S^{-4/5} \ell_p}.
\label{alpha7b}
\end{equation}

\begin{figure}[t]
\begin{tabular}{lr}
\includegraphics[scale=0.55]{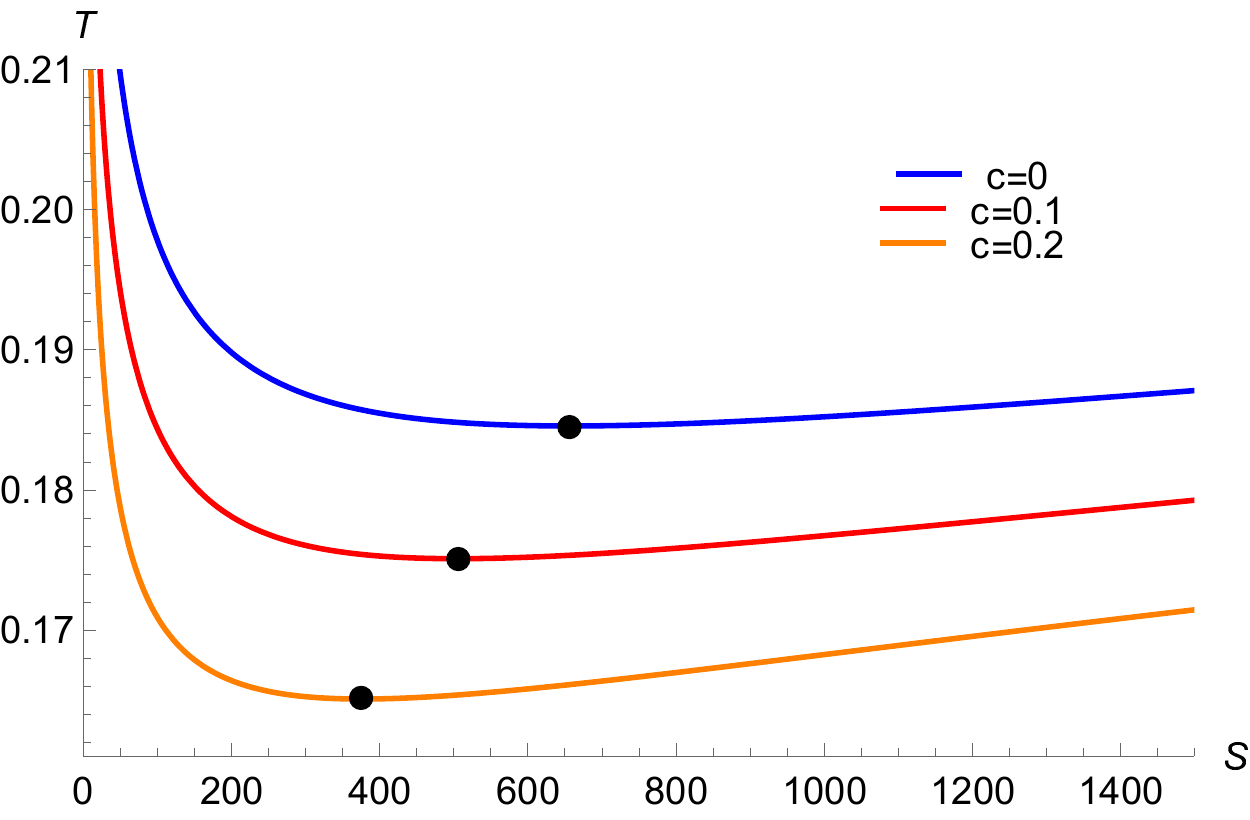}&\hspace{0.4cm}
\includegraphics[scale=0.40]{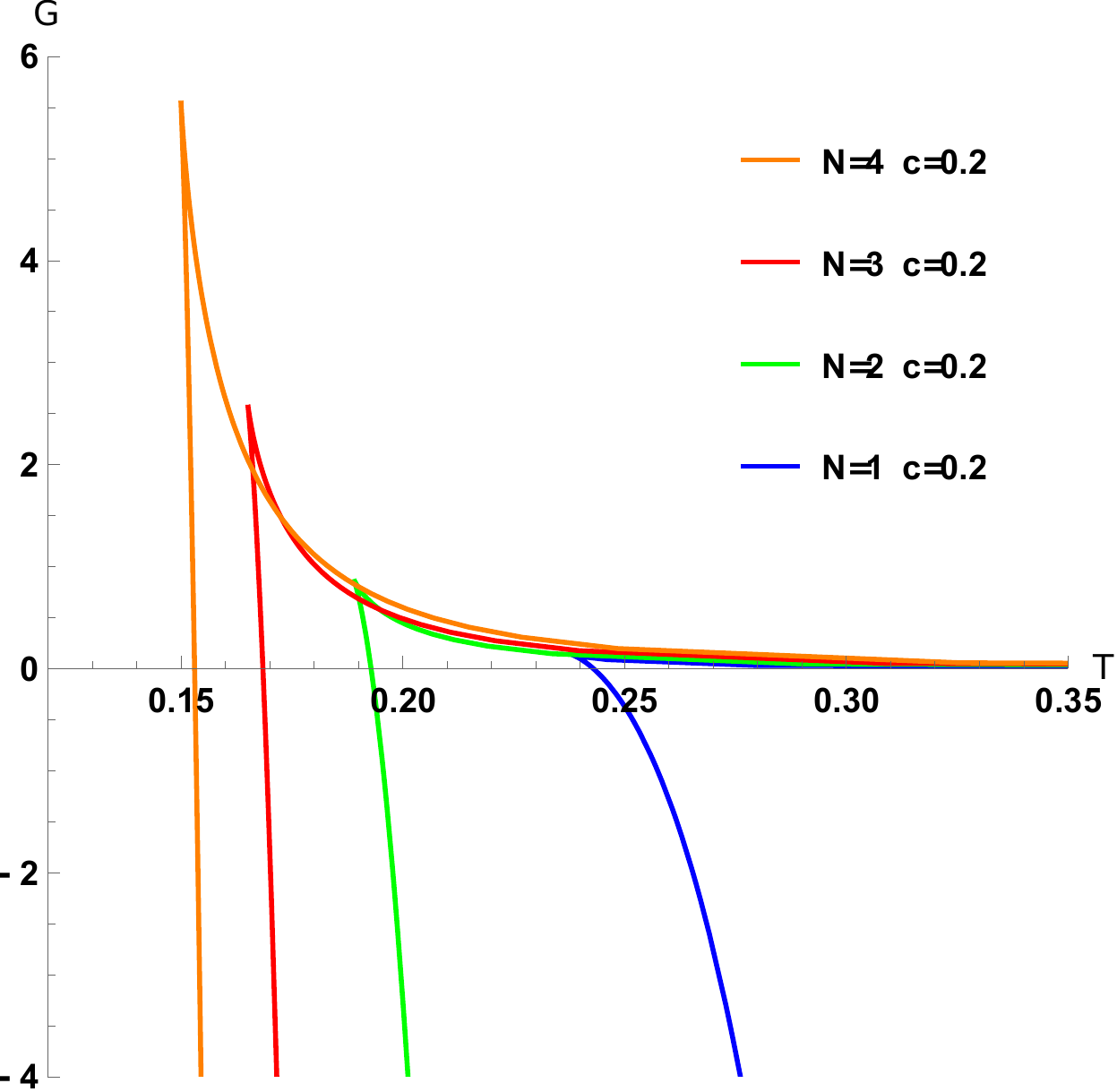}
\end{tabular}
\caption{ (Left) Temperature as function of the entropy using $N=3$ and $\ell_{p}=1$ and different values of $c$. The dots 
denote the minimum of the temperature for each case.
(Right) The Gibbs free energy as function of temperature for different values of $N$, $\ell_{p}=1$ and $c=0.2$. The sign of Gibbs free energy changes at  
the Hawking-Page temperature $T_7^{HP}$.}
\label{G7NDE}
\end{figure}

To investigate the associated phase transitions behaviours, we first plot the Hawking temperature as a function of the entropy as illustrated in figure \ref{G7NDE} for different values of the intensity $c$.

For a general number of $M5$-branes and a general intensity of DE dynamical field, we can show that the Hawking temperature has a minimum for
\begin{equation}
S_7^{DE-min}=\frac{2^{17/2}}{3^{7/2}} \, \left( 1 -c \right)^{5/2} \pi N^{3}.
\end{equation}
This corresponds to the following minimal temperature
\begin{equation}
T_7^{DE-min}=\frac{3^{1/2} \sqrt{1-c}}{2^{1/2} \, \pi ^{4/3} N^{1/3} \ell_{p}}.
\end{equation}
From figure \ref{G7NDE}, we see that the temperature is decreasing when the DE intensity $c$ increases. This shows that DE behaves like a cooling system surrounding the black hole. This confirms the results obtained in \cite{notrea}. The minimum of the temperature and the corresponding entropy noted by the black dotes in such a figure are also affected in the same way.

For more investigations in the phase transitions of the seven-dimensional $AdS$ black hole,  we illustrate the Gibbs free energy as a function of the Hawking temperature $T_7^{(-3) \, DE}$ for different values $N$ of $M5$-branes and a fixed value of $c$. This is given  in figure \ref{G7NDE}.

\begin{figure} 
\begin{center}
\includegraphics[scale=0.4]{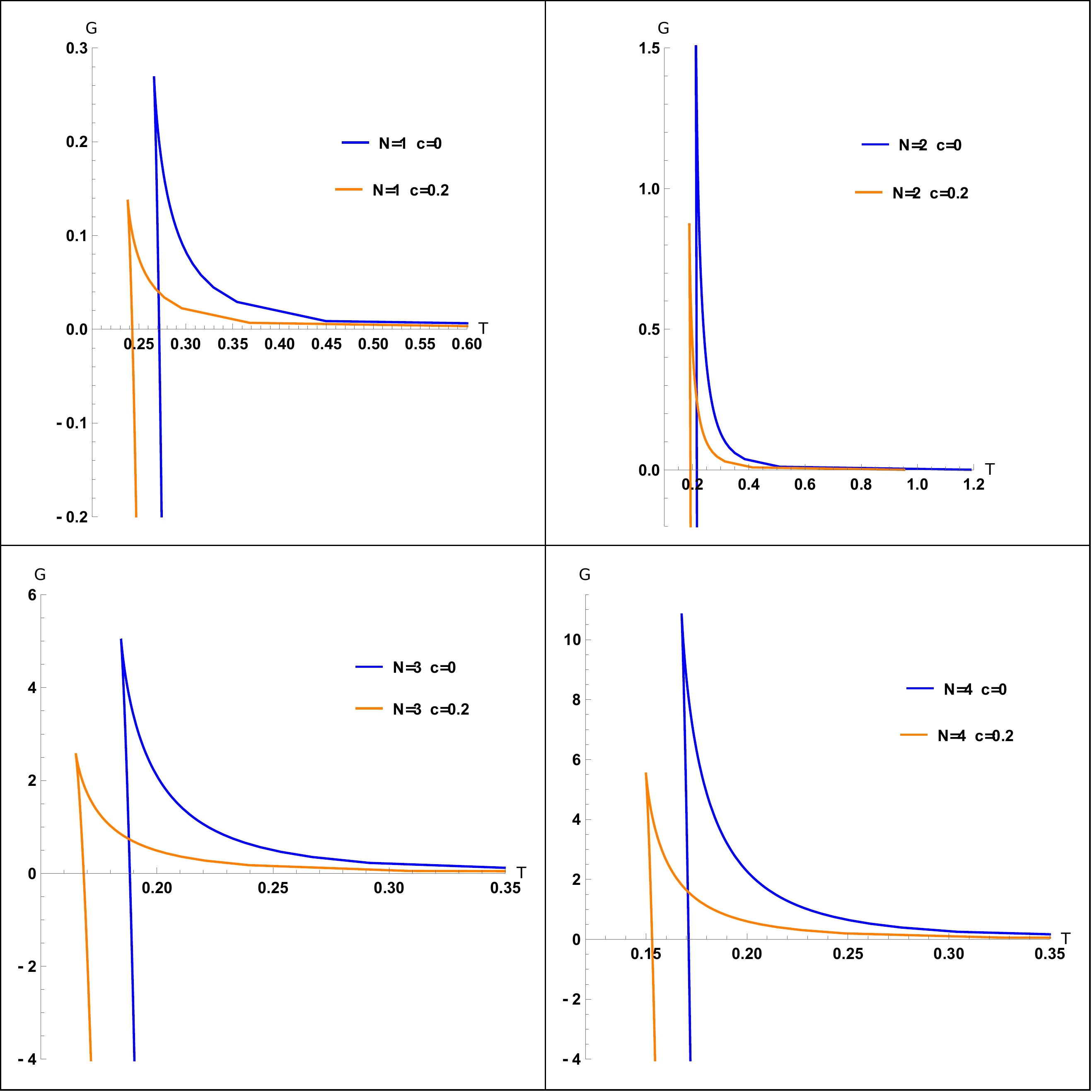}
\caption{The Gibbs free energy as function of the temperature in the absence of DE and in its presence for different values of $N$, $\ell_{p}=1$.}
\label{G7DE}
\end{center}
\end{figure}

It is remarked, from this figure,  that the phases discussed in figure \ref{G7NDE} are getting smaller. The Gibbs free energy decreases also, yielding a smaller unstable black hole phase.
From the Gibbs free energy given in Eq.\eqref{GDE70}, one can find the Hawking-Page phase transition 
corresponding to
\begin{equation}
S_7^{DE-HP}=\frac{2^{6}}{3} \left( 1-c \right)^{5/2} \pi N^{3}.
\end{equation}
Thus, the Hawking-Page temperature is
\begin{equation}
T_7^{DE-HP}=\frac{5\sqrt{1-c}}{4 \, \pi ^{4/3} N^{1/3} \ell_{p}},
\end{equation}
which is smaller than $T_7^{DE-min}$.
To identify the difference in the behavior of the Gibbs free energy in the presence of quintessence, we plot this function for $c=0$ (absence of DE) and for $c=0.2$ for several values $N$  of $M5$-branes in figure \ref{G7DE}.

We see that the decrease in the radiation phase and the phase where no black hole can exist comes from the temperature diminution. However, the stable and the unstable phases are directly affected by the diminution of the Gibbs free energy.

\begin{figure}[t]
\centering
\begin{tabular}{l}
\includegraphics[scale=0.55]{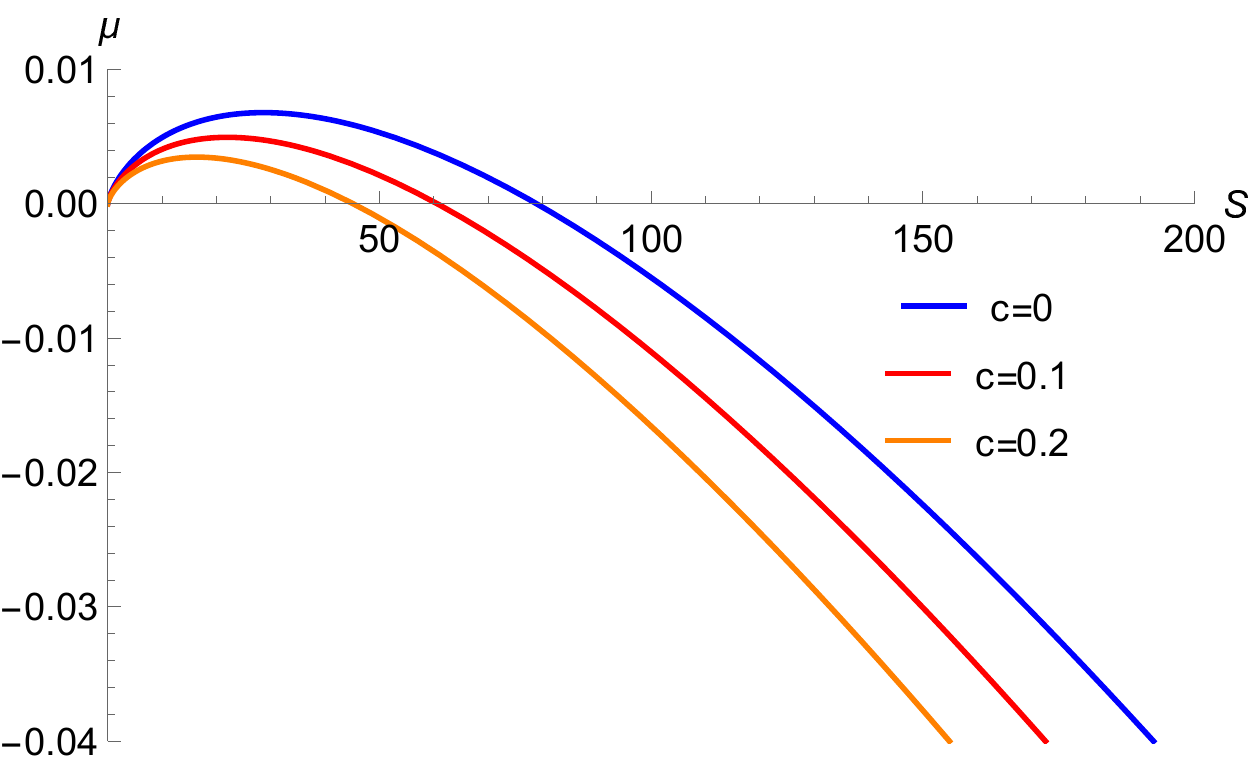}\\
\includegraphics[scale=0.65]{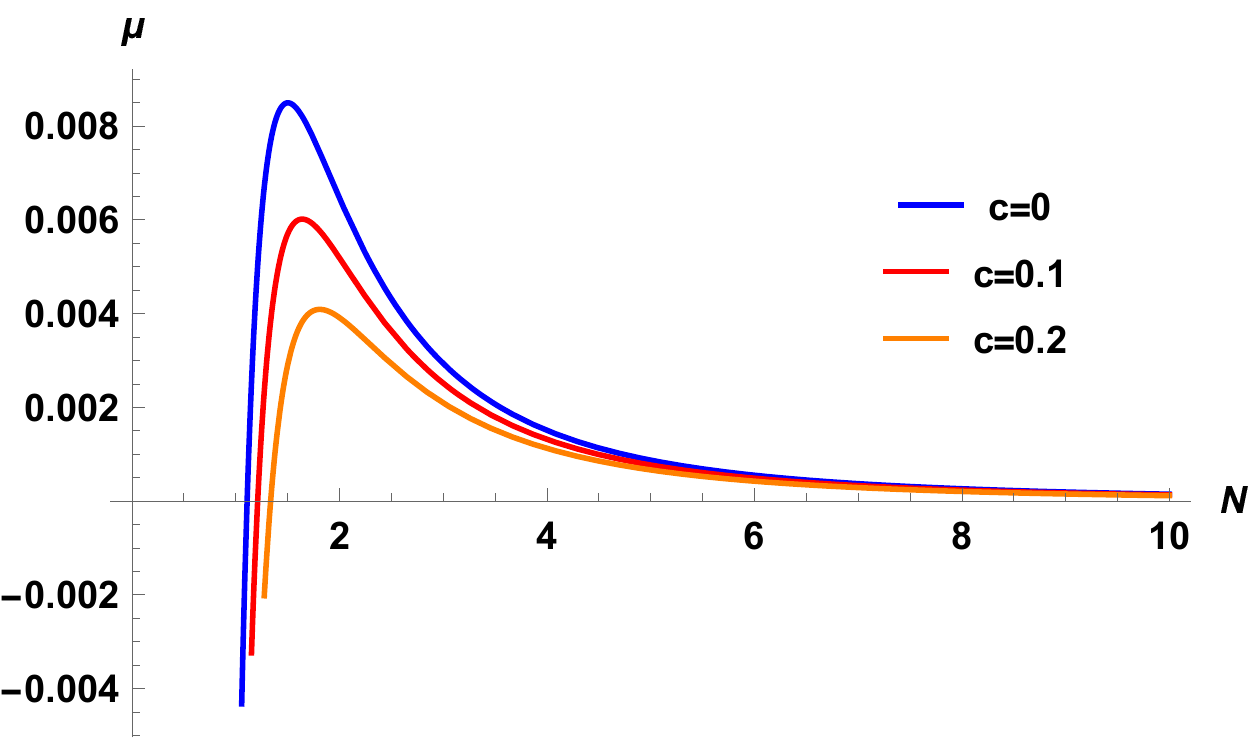}\\
\includegraphics[scale=0.45]{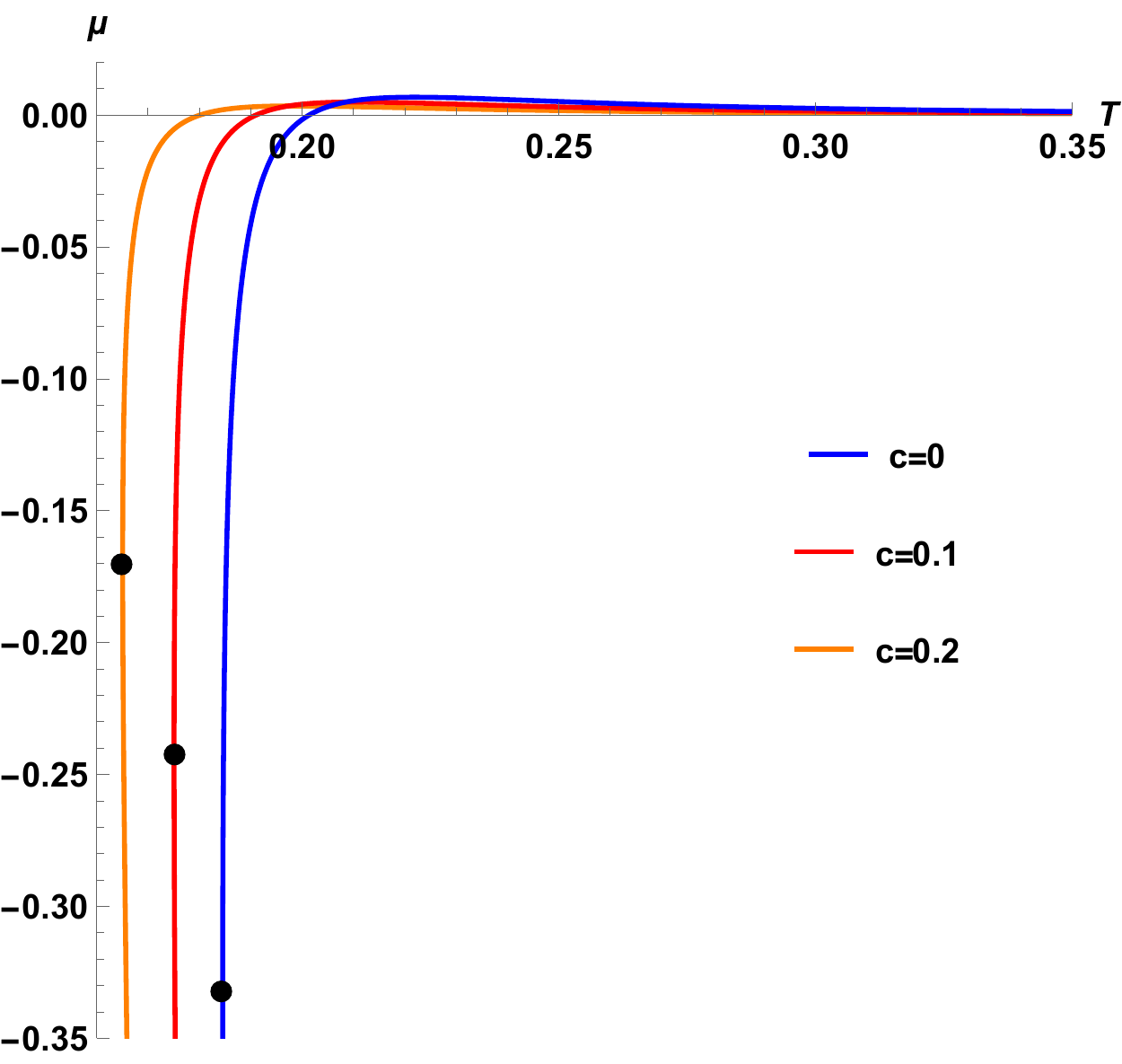}
\end{tabular}
\caption{(Top) The chemical potential as a function of entropy for $N = 3, \ell_{p}=1$. The sign change
 of the chemical potential happens at $S_7^{DE-\mu=0}$ .
(Center) The chemical potential as a function of the number of $M5$-branes $N$. Here we take $\ell_{p} = 1$ and $S_7^{max} = 4$. 
(Bottom) The chemical potential as a function of temperature $T$. We take $\ell_{p} = 1$ and $N = 3$.}
\label{mu7NDE}
\end{figure}

It has been observed that the quintessence field affects also the chemical potential. To inspect the corresponding modifications, we plot the chemical potential as a function of entropy $S$ for $N=3$ in figure \ref{mu7NDE}.

It is remarked that the chemical potential is positive for small entropy $S$, and negative for large
$S$. Moreover, the figure \ref{mu7NDE} shows a diminution of both the chemical potential and the entropy when DE is present. Besides, the gap between the curves seems to increase when the entropy increases.
The chemical potential changes the sign at the following entropy
\begin{equation}
S_7^{DE-\mu=0}=\frac{2^{17/2}}{3 \cdot 7^{5/2}} \,\left(1-c \right)^{5/2} \pi N^3.
\end{equation}
As mentioned above, we can show that $S_7^{DE-\mu=0}<S_7^{DE-min}<S_7^{DE-HP}$.

Furthermore, we can also study the behavior of the chemical potential as a function of the number $N$ of $M5$-branes in such a  M-theory compactification. This can be illustrated in figure \ref{mu7NDE}. Here, the entropy $S$ has a fixed value. The maximum of the chemical potential corresponds to the point
\begin{equation}
S_7^{DE-max} = \frac{2^{17/2}}{3 \cdot 7^{5/2}} \, \left( \frac{41}{59} \right)^{5/2} \, \left( 1-c \right)^{5/2} \pi \left( N_7^{DE-max} \right) ^3,
\end{equation}
namely, $N_7^{DE-max} \simeq 1,81$ for $S_7^{max}=4$ and $c=0.2$. From $N_7^{DE-max}$, we see that the number of $M5$-branes grows in the presence of DE in the M-theory compactification on $\mathbb{S}^4$.

Besides, we can also reveal that DE stabilises the $AdS$ black hole. In figure \ref{mu7NDE}, we plot the chemical potential as a function of temperature $T_7^{ (-3) \, DE}$ for a fixed $N$, where the dots denotes $T_7^{DE-min}$. Lower from this point, we have $T_7^{DE-HP}$ which separates the lower stable branch and the upper unstable branch where $T_7^{DE-min}$ resides.

We notice that $T_7^{DE-min}$ is getting higher in the curves when the intensity of DE $c$ is bigger, which is also true for $T_7^{DE-HP}$. In the presence of DE, then, the unstable branch is getting smaller. However, the stable one becomes relevant.

%
%

\subsection{DE effect on $AdS$ black holes }

Let us have a closer  look on the effects of DE energy presence on $d$-dimensional $AdS$ black holes embedded in 
M-theory/superstring inspired models. A close examination shows that the $d$-dimensional $AdS$ black hole entropy 
can be put in a compact formula given by
\begin{equation}
S_d^{DE-i}\left(N,c\right)=\left( 1-c \right)^{\frac{d-2}{2}} \cdot S_d^i\left(N\right),
\label{Sgen}
\end{equation}
where $i$ stands for the set $\{ min, HP\}$. 
It is worth noting that $(1-c)$ is interpreted as a DE scaling factor depending on the dimension $d$ of the 
considered $AdS$ black hole. We observe a decrease in the entropy values of such $AdS$ black holes. This means that 
DE induces the reduction of  the associated number of microstates.

Putting this entropy in the equation of the Hawking temperature given in \eqref{Tgf}, the associated $T_d^{DE-i}$ temperature 
can be obtained. Indeed, it satisfies the following general formula
\begin{equation}
T_d^{DE-i}\left(N,c\right)=\left( 1-c \right)^{\frac{1}{2}} \cdot T_d^i\left(N\right).
\label{Tgen}
\end{equation}
For the  temperature, however, DE scaling factor does not depend on the dimension of the $AdS$ black holes. This can be understood that the temperature does not depend on the size parameter of the theory in question. It follows also that Eq.\eqref{Tgen} provides a colder black hole, being a  more stable one.
Moreover, we observe an increasing behavior regarding the number $N$ of $(d-2)$-branes. In $d$ dimensions, such a number takes the following general form
\begin{equation}
N_d^{DE-max}\left(S,c\right)=\frac{N_d^{max}\left(S\right)}{\left( 1-c \right)^{\frac{d-2}{d-1}}}.
\end{equation}
In M-theory/superstring inspired models, it follows that $N_d^{DE-max}\left(S,c\right)$ grows with the DE contributions. Thus, DE enhances the number of branes by generating non trivial extra branes which we refer to as \textit{"Dark-branes"}.

\section{Conclusion and open questions}

In this work, we    have  investigated the thermodynamical phase transitions of $d$-dimensional $AdS$ black holes surrounded by DE. These black hole solutions  have been   embedded in $D$-dimensional superstring/M-theory inspired models with the  $AdS_d \times \mathbb{S}^{d+k}$ space-time, where $D=2d+k$ is their Minkowski dimension.   These models, which  could be associated with $N$ coincident $(d-2)$-branes supposed to live in such higher dimensional inspired  theories, have been labeled by a triplet $(D,d,k)$ where $k$   carries data  on   the internal space $ \mathbb{S}^{d+k}$.   By interpreting the cosmological constant as the number of colors $N^{\frac{d-1}{2}}$, we have  computed various thermodynamical quantities denoted by $X^{(k) DE}_d$  in terms of the brane number $N$, the entropy $S$ and  DE contributions   via  a dynamical quintessence scalar   field.  By calculating  the chemical potential conjugated to the number of colors in the absence of DE, we have found that the black hole is more stable,  enjoying  a large number of branes for lower dimensions $d$.
In the presence of DE, we have realised that the state parameter $\omega_q$ should have specific values, for $(D,d,k)$ models,  providing non trivial phase  transition results.

\begin{table}[h]
\begin{center}
\begin{tabular}{|l||l|}
\hline
$AdS_{d}\times \mathbb{S}^{d+k}$ & $w_q$\\
\hline
\hline
$AdS_{4}\times \mathbb{S}^{7}$ &$- \frac{1}{3}$ \\
\hline
$AdS_{5}\times \mathbb{S}^{5}$ &$- \frac{1}{2}$ \\
\hline
$AdS_{7}\times \mathbb{S}^{4}$ & $-\frac{2}{3} $\\
\hline
\end{tabular}
\caption{Summary of results of different cases for $\omega_q$.}
\label{Tableau3}
\end{center}
\end{table}

 Among others, we have obtained a smaller \textit{no black hole} phase, a more stable and colder black hole. Furthermore,  we have  found  an enhancement regarding the number of branes which we refer to as \textit{Dark-branes}. We believe that such suggestions need deeper investigations. We hope to come back to this non trivial remark in connection with cosmology in future works.

Inspired by sphere compactifications in higher dimensional theories, various models could be examined for quintessential $AdS$ black holes. A possible  situation  is associated with  trivial sphere fibrations  with the  same dimension $n$. In this way, the internal space $X^{d+k}$ can be factorized as
\begin{equation}
X^{d+k}= \mathbb{S}^{n} \times \mathbb{S}^{n} \times \cdots \times \mathbb{S}^{n}.
\end{equation}
 Borrowing ideas from intersecting attractors \cite{2}, these models could provide non trivial phase transitions corresponding to such geometric fibrations. In this context, lower dimensional cases could be approached using group theory techniques. Another road is to think about orbifolding spheres generating twisted sectors in the resulting compactified theories. This could bring new features to $AdS$ black holes surrounded by non trivial  contributions including dark matter. Once these sphere geometries become accessible, Calabi-Yau manifolds  could  find places in the building of  $AdS$ black holes from such  M-theory/superstring inspired models.

\vspace{0.5cm}
\section*{Acknowledgments}

 AB would
like to thank the Departamento de F\'isica, Universidad de Murcia for very kind hospitality and scientific supports 
during the realization of a part of this work and  he thanks  J. J. Fern\'andez-Melgarejo,  H. El Moumni, M. B. Sedra and A. Segui for  discussions on related topics.  The work of ET  has been  supported in part by
the Spanish Ministerio de Universidades and Fundacion
Seneca (CARM Murcia) grants FIS2015-3454, PI2019-
2356B and the Universidad de Murcia project E024-018. This work is partially
supported by the ICTP through AF-13.

\appendix

\section{Eleven-dimensional compactification: The $AdS_{4}\times \mathbb{S}^7$ case without DE}

\begin{figure} 
\begin{center}
\includegraphics[scale=0.55]{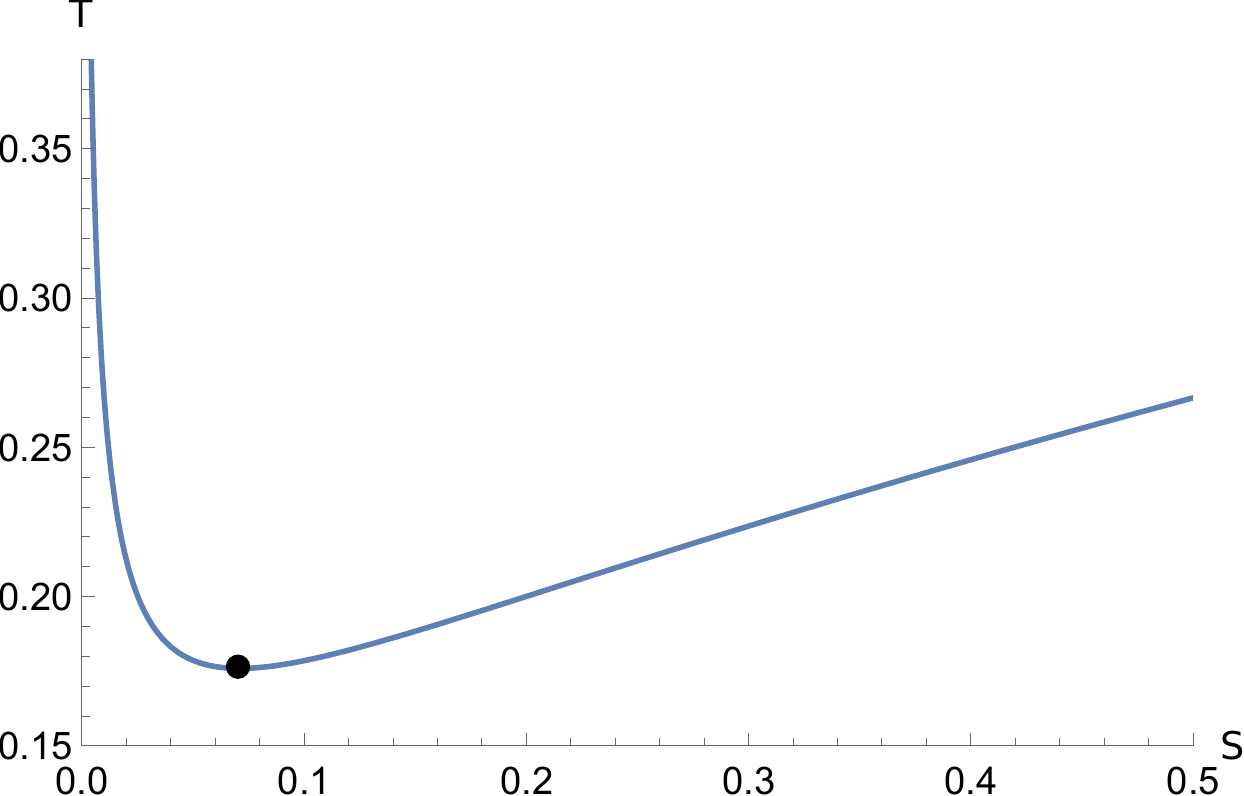}
\caption{Temperature as function of the entropy using $N=3$ and $\ell_{p}=1$.}
\label{T4}
\end{center}
\end{figure}

We consider  here 
 eleven-dimensional M-theory, with $ AdS_{4}\times \mathbb{S}^{7}$ space-time in the presence of $N$ coincident $M2$-branes, i.e.  a model with characterised by
the triplet
\begin{equation}
(D,d,k)=(11,4,3).
\end{equation}
Using the same procedure  as in  previous sections, we  compute the thermodynamical quantities $X_4^{(3)}$ for such four-dimensional $AdS$ black holes obtained from the compactification of M-theory on seven dimensional real sphere $\mathbb{S}^{7}$. In this case 
 the volume factor is
\begin{equation}
\omega_{7}=\frac{ \pi^{4}}{3}.
\label{omegaAds4}
\end{equation}
 We list  the final expressions of some of these quantities:
\begin{itemize}
\item the mass parameters

\begin{equation}
M_{4}^{(3)}(S,N)=\frac{ \pi^{3/2} \,  S^{\frac{1}{2}} \, N^{\frac{7}{12}} + 96 \sqrt{2} \times S^{\frac{3}{2}} N^{\frac{-11}{12}} }{8 \times   2^{7/12} \, \sqrt{3} \,  \pi^{25/12} \, \ell_{p}},
\label{M40}
\end{equation}

\item the temperature
\begin{equation}
T_{4}^{(3)}(S,N)=\frac{ \pi^{3/2} \,  S^{-\frac{1}{2}} \, N^{\frac{7}{12}} +3 \times 96 \sqrt{2} \, S^{\frac{1}{2}} N^{\frac{-11}{12}} }{16 \times   2^{7/12} \, \sqrt{3} \,  \pi^{25/12} \, \ell_{p}},
\label{T40}
\end{equation}

\item chemical potential
\begin{equation}
\mu_{4}^{(3)}(S,N)=\frac{ 7 \times \pi^{3/2} \,  S^{\frac{1}{2}} \, N^{-\frac{5}{12}} -11 \times 96 \sqrt{2} \, S^{\frac{3}{2}} N^{\frac{-23}{12}} }{96 \times   2^{7/12} \, \sqrt{3} \,  \pi^{25/12} \, \ell_{p}},
\label{mu40}
\end{equation}
\item Gibbs free energy
\begin{equation}
G_{4}^{(3)}(S,N)=\frac{ \pi^{3/2} \,  S^{\frac{1}{2}} \, N^{\frac{7}{12}} - 96 \sqrt{2} \times S^{\frac{3}{2}} N^{\frac{-11}{12}} }{16 \times   2^{7/12} \, \sqrt{3} \,  \pi^{25/12} \, \ell_{p}}.
\label{G40}
\end{equation}
\end{itemize}
To investigate the existence of phase transitions, 
we first consider  the behavior of the Hawking temperature with respect to entropy. This is illustrated in figure \ref{T4}.

\begin{figure} 
\begin{center}
\includegraphics[scale=0.5]{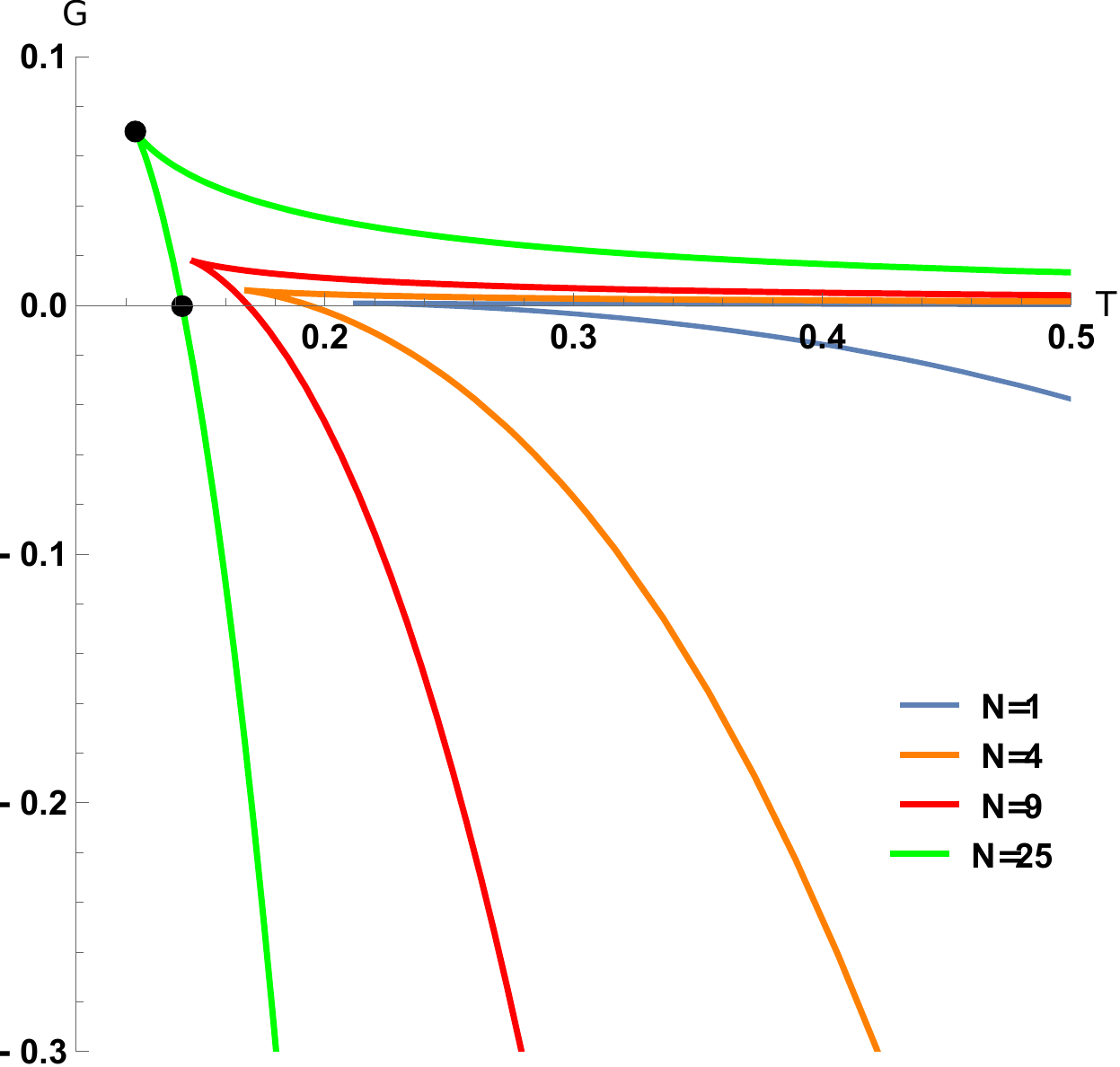}
\caption{The Gibbs free energy as function of temperature for different values of $N$ and $\ell_{p}=1$.}
\label{G4N}
\end{center}
\end{figure}

The temperature reaches its minimum at $S_{4}^{min} \simeq 0.07$. For a general number of $M2$-branes $N$, one can reveal that the Hawking temperature has a minimum at
\begin{equation}
S_{4}^{min}=\frac{\pi^{3/2}}{3^{2} \, \cdot \, 2^{11/2}} N^{3/2},
\end{equation}
corresponding to the minimal temperature
\begin{equation}
T_{4}^{min}=\frac{3^{1/2}}{2^{5/6} \, \pi^{4/3} N^{1/6} \ell_{p}}.
\end{equation}
Under the minimal temperature no black hole solution can survive. However, we observe from figure \ref{T4} that above such a temperature two branches arise. The first one is associated with small entropy corresponding to a thermodynamically unstable black hole. The second one, with large entropy, indicates a thermodynamically stable black hole.\\
To discuss the brane number $N$ effect on the phase transition, we use the computed Gibbs free energy. In figure \ref{G4N}, we plot the Gibbs free energy as a function of the Hawking temperature $T_4^{(3)}$ for different values $N$ of $M2$-branes in  such M-theory backgrounds.

As before the sign of Gibbs free energy changes at the Hawking-Page temperature $T_4^{HP}$ associated with $S_4^{HP}$, corresponding to the Hawking-Page phase transition and given by the lower dot in figure \ref{G4N}. From the Gibbs free energy given in Eq.\eqref{G40}, such a phase transition appears at
\begin{equation}
S_{4}^{HP}=\frac{ \pi^{3/2} N^{3/2}}{3 \cdot 2^{11/2}}.
\end{equation}
The corresponding Hawking-Page temperature is
\begin{equation}
T_4^{HP}=\frac{2^{1/6}}{\pi ^{4/3} N^{1/6} \ell_{p}},
\end{equation}
being smaller than $T_{4}^{min}$. 

In figure \ref{G4N}, a first order Hawking-Page phase transition arises at $T_4^{HP}$ between large (stable) black holes and the thermal radiations. Besides, one can see that below the higher dot representing $T_4^{min}$ no black hole can survive. For lower Gibbs free energy, we have two branches. The upper branch describes a small (unstable) black hole with a negative specific heat. However, the lower one indicates a (large) stable black hole solution with a positive specific heat.

\begin{figure} 
\begin{center}
\includegraphics[scale=0.55]{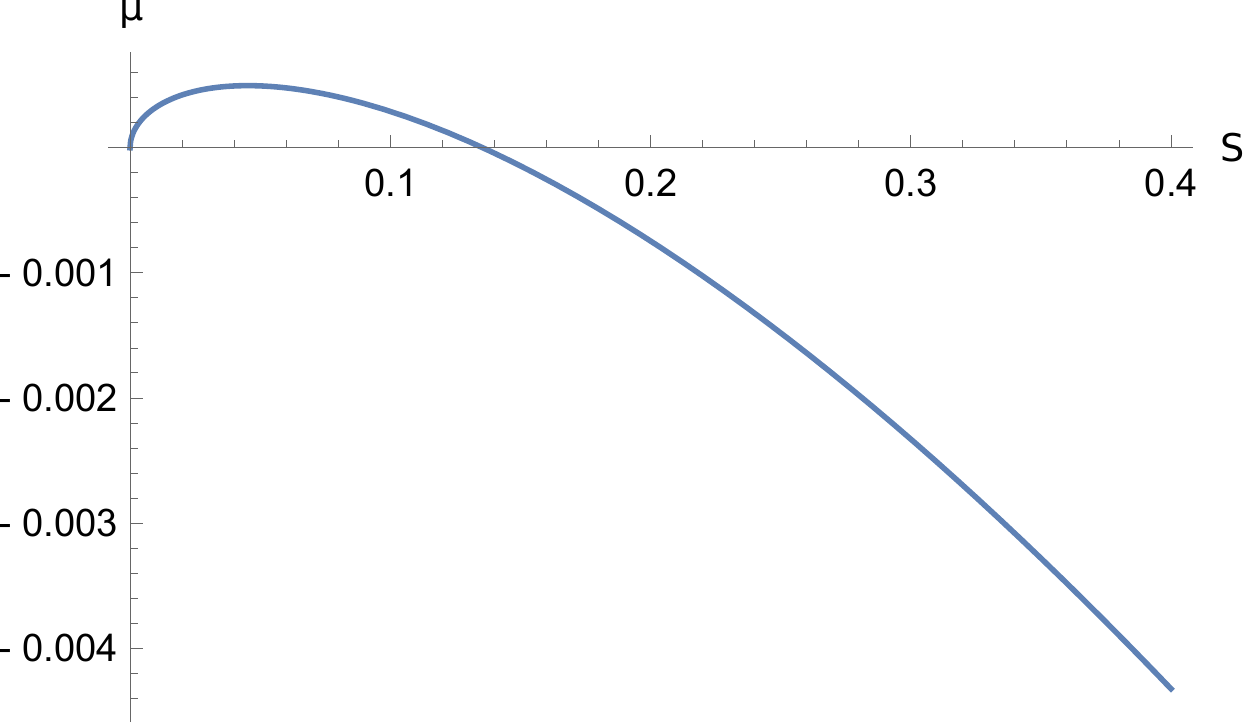}
\caption{The chemical potential as a function of entropy for $N = 3, \ell_{p}=1$. The sign  change  of the chemical potential happens at $S_{4}^{\mu=0} \simeq 0.13$.}
\label{mu4}
\end{center}
\end{figure}

The chemical potential can be also exploited to examine the phase transitions of the studied four-dimensional AdS black hole embedded in the M-theory compactification on $\mathbb{S}^{7}$. In figure \ref{mu4}, we plot the chemical potential as a function of the entropy $S$ for a fixed number of $M2$-branes. For simplicity reasons, we consider the case associated with $N=3$.

From such a figure, we remark that the chemical potential is positive for small entropy $S$, and negative for large
entropy $S$. It has been found that the sign change in the chemical potential occurs at the following entropy
\begin{equation}
S_{4}^{\mu=0}=\frac{7 \, \pi ^{3/2} }{11 \times 3 \cdot 2^{11/2}} N^{3/2}.
\label{s4mu0}
\end{equation}
In this case, the four-dimensional $AdS$ black hole verifies $S_{4}^{min}<S_{4}^{\mu=0}<S_{4}^{HP}$. We remark that $S_{4}^{min}$ is the lowest value of the entropy. 

\begin{figure} 
\begin{center}
\includegraphics[scale=0.55]{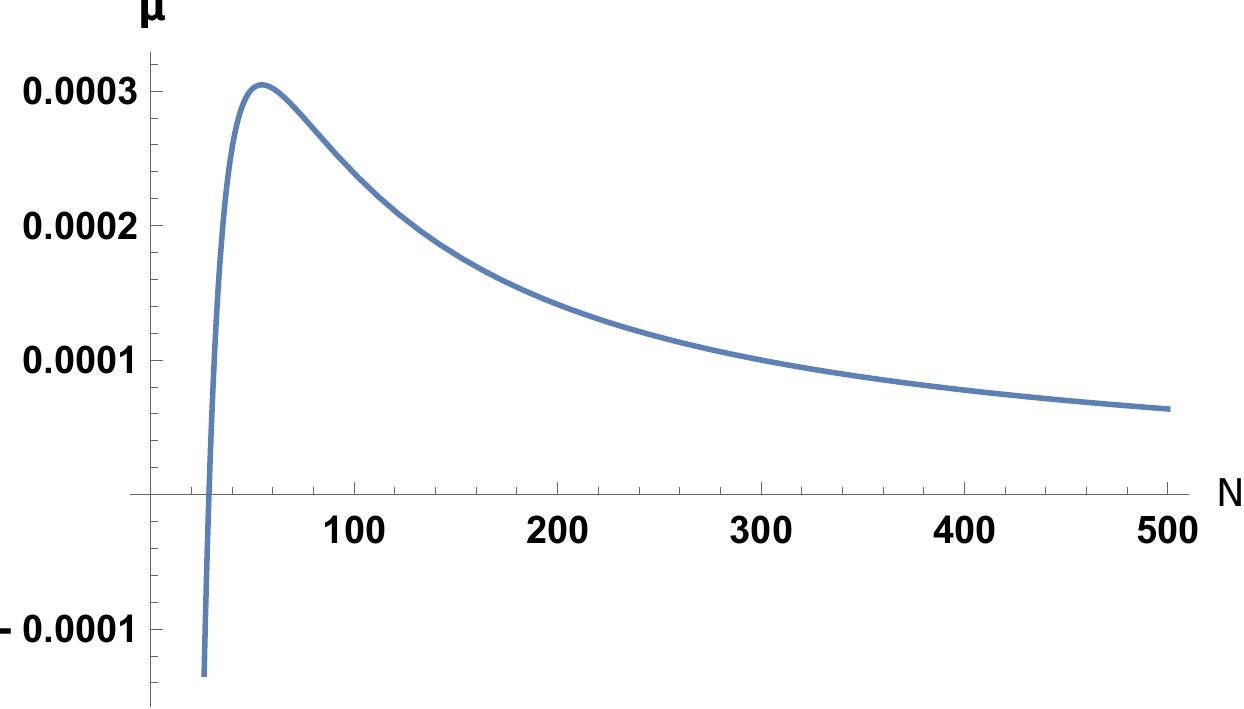}
\caption{The chemical potential as a function of the number of $M2$-branes $N$. Here we take $\ell_{p} = 1$ and $S = 4$. }
\label{mu4N}
\end{center}
\end{figure}

In terms of the  temperature, the chemical potential changes its sign at the following thermal point
\begin{equation}
T_4^{\mu=0}=\frac{8 \, 2^{1/6}}{\sqrt{77} \pi^{4/3} \ell_p N^{1/6}}.
\label{t4mu0}
\end{equation}
In figure \ref{mu4N}, the chemical potential is plotted as a function of $N$ with a fixed entropy $S$.

\begin{figure} 
\begin{center}
\includegraphics[scale=0.55]{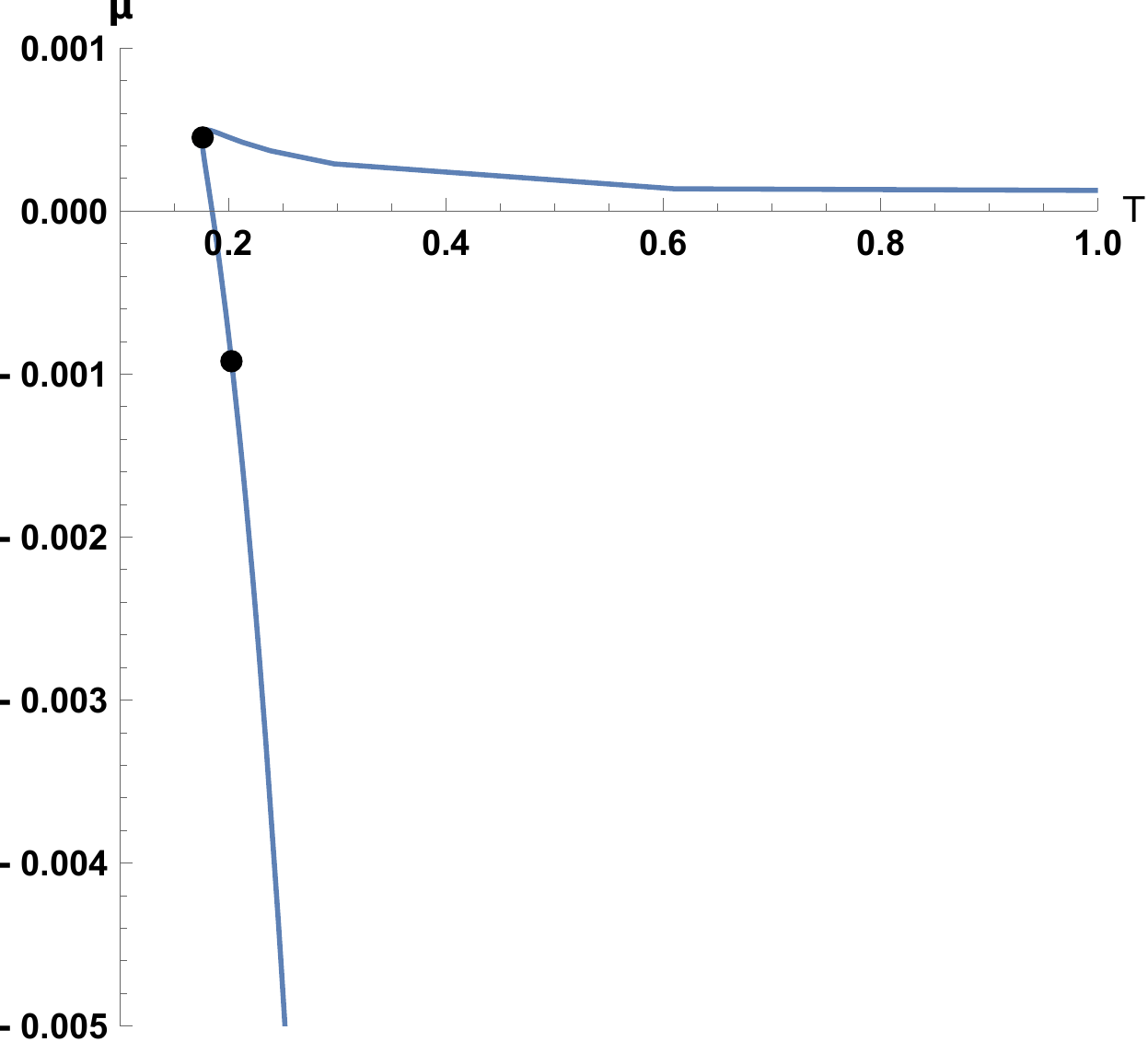}
\caption{The chemical potential as a function of temperature $T$. We take $\ell_{p} = 1$ and $N = 9$.}
\label{mu4T}
\end{center}
\end{figure}

From figure \ref{mu4N}, it is observed that the maximum of the chemical potential is associated with the value $ S_{4}^{max} = \frac{7 \times \, \pi^{3/2}}{3 \times 29\, \cdot 2^{11/2}} \left( N_{4}^{max} \right)^{3/2} $, namely, $N_{4}^{max} \simeq 54 $ for $S_{4}^{max}=4$.\\
In figure \ref{mu4T}, we plot the chemical potential as a function of the temperature $T_4^{(3)}$ for a fixed number $N$ of $M2$-branes.

It is noted that the upper dot in figure \ref{mu4T} indicates the minimum of temperature 
$T^{min}_4$. Bellow this point, however, one can find the Hawking-Page temperature that separates the lower  stable branch  and the upper unstable branch where $T_4^{min}$ resides. Here, the chemical potential changes the sign at unphysical regions. From Eq.\eqref{s4mu0}, the
chemical potential is positive when the number of $M2$-branes is constrained by $N^{3/2} > \frac{11 \times 3 \cdot 2^{11/2} }{7} \, \frac{S}
{\pi^{3/2}}$. This limit can be saturated for
\begin{equation}
T^{\mu=0}_4 \simeq 0.9\, T_4^{HP}.
\end{equation}
We remark that $T_4^{\mu=0} < T_4^{HP}$ implying that the pure $AdS$ geometry is preferred over the black hole.

\section{Ten-dimensional compactification: The $AdS_{5}\times \mathbb{S}^{5}$ case without DE}

\begin{figure} 
\begin{center}
\includegraphics[scale=0.55]{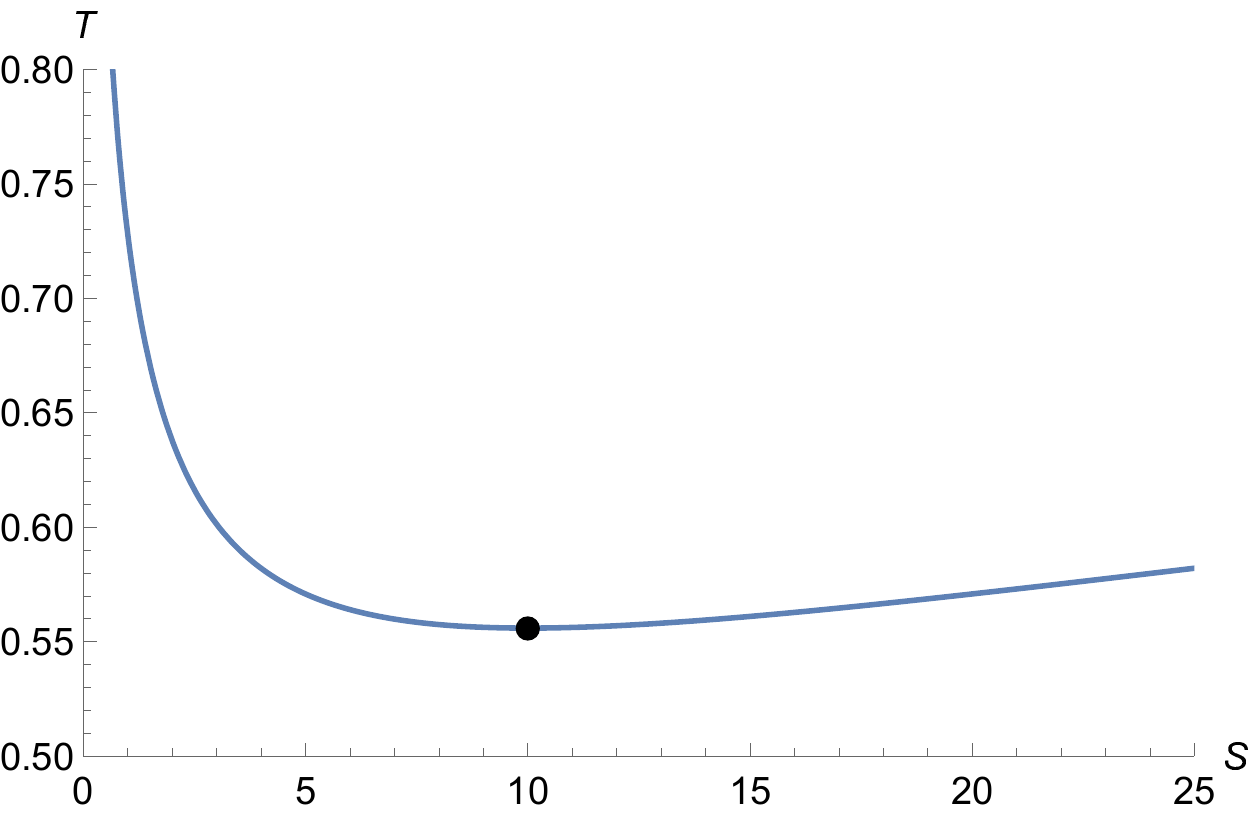}
\caption{Temperature as function of the entropy using $N=3$ and $\ell_{p}=1$. }
\label{T5}
\end{center}
\end{figure}

In this part, we reconsider the case of ten-dimensional compactification associated with the odd-dimensional spheres. In our formulation, the model that we deal with is indexed by
\begin{equation}
(D,d,k)=(10,5,0).
\end{equation}
This case is nothing but the ten-dimensional type IIB superstring theory with the $AdS_{5}\times \mathbb{S}^{5}$ space-time in the presence of $D3$-branes. The corresponding five-dimensional $AdS$ black hole is obtained from the compactification on  the odd dimensional real sphere $ \mathbb{S}^{5}$ with the following volume factor
\begin{equation}
\omega_{5}=\pi^{3}.
\label{omegaAds5}
\end{equation}
In this way, the gravitational constant is $G_{10}= \ell_{p}^{8}$. Taking into account Eq.\eqref{omegaAds5}, we get
\begin{itemize}
\item mass
\begin{equation}
M_{5}^{(0)}(S,N)=\frac{ 3 \left( \pi^{2/3} \,  S^{\frac{2}{3}} \, N^{\frac{5}{12}} + S^{\frac{4}{3}} N^{-\frac{11}{12}} \right)}{4 \times   2^{1/8} \,  \pi^{5/6} \, \ell_{p}},
\label{M50}
\end{equation}
\item temperature
\begin{equation}
T_{5}^{(0)}(S,N)=\frac{ \pi^{2/3} \,  S^{-\frac{1}{3}} \, N^{\frac{5}{12}} + 2 \times S^{\frac{1}{3}} N^{-\frac{11}{12}} }{2 \times   2^{1/8} \,  \pi^{5/6} \, \ell_{p}},
\label{T50}
\end{equation}

\item chemical potential
\begin{equation}
\mu_{5}^{(0)}(S,N)=\frac{ 5 \times \pi^{2/3} \,  S^{\frac{2}{3}} \, N^{-\frac{19}{12}} - 11 \times S^{\frac{4}{3}} \,  N^{-\frac{35}{12}} }{32 \times   2^{1/8} \,  \pi^{5/6} \, \ell_{p}},
\label{mu50}
\end{equation}
\item Gibbs free energy
\begin{equation}
G_{5}^{(0)}(S,N)=\frac{ \pi^{2/3} \,  S^{\frac{2}{3}} \, N^{\frac{5}{12}} - S^{\frac{4}{3}} N^{-\frac{11}{12}} }{4 \times   2^{1/8} \,  \pi^{5/6} \, \ell_{p}}.
\label{G50}
\end{equation}
\end{itemize}
The obtained results match perfectly with the ones given in \cite{I28}. However, we reconsider and refine the associated results by presenting detailed discussions. This will be needed later on when we turn on the DE effects.\\
To discuss the associated phase transitions, we first plot the behavior of the Hawking temperature with respect to the entropy. This is illustrated in figure \ref{T5}.

\begin{figure} 
\begin{center}
\includegraphics[scale=0.5]{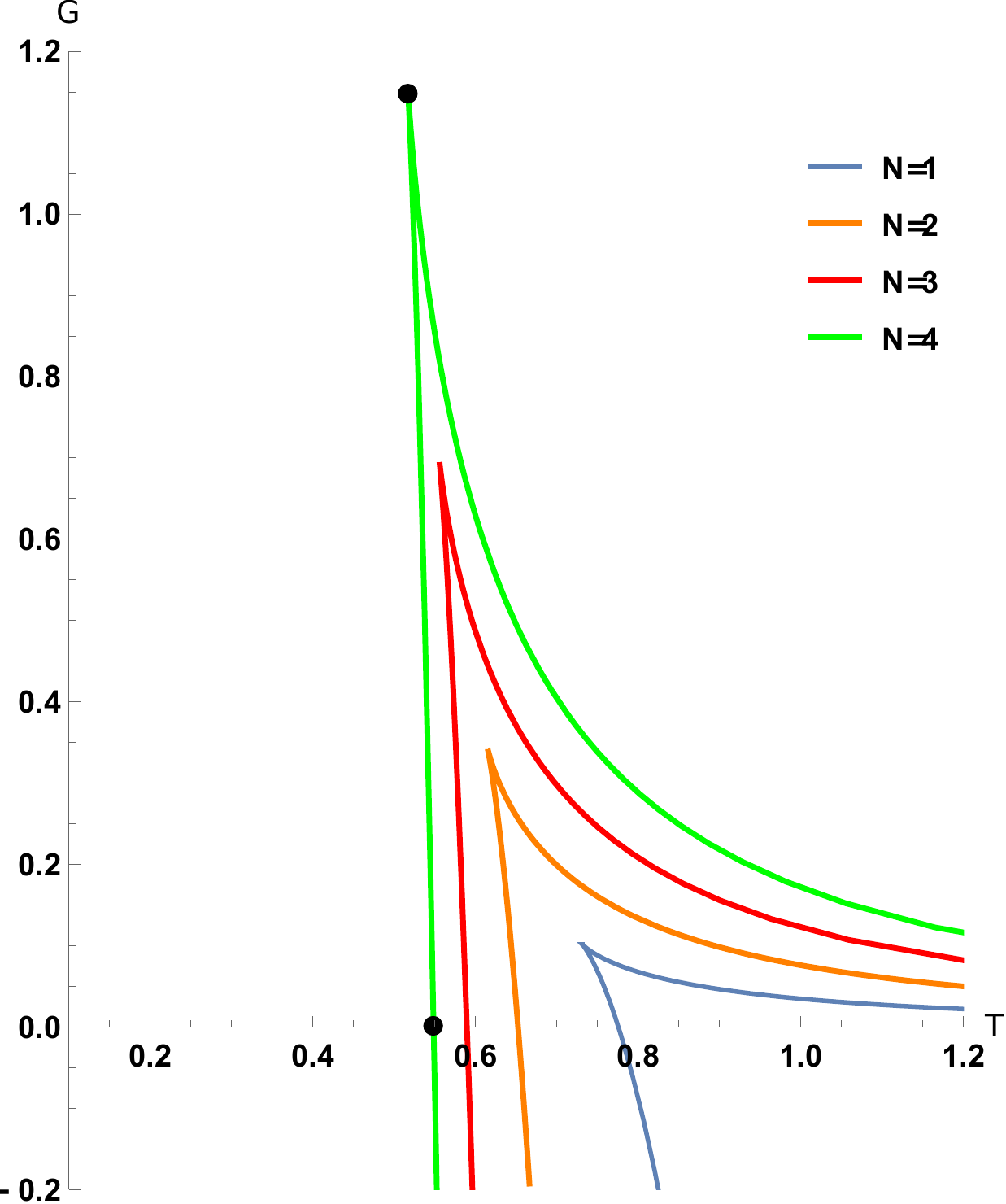}
\caption{The Gibbs free energy as function of temperature for different values of $N$ and $\ell_{p}=1$.}
\label{G5N}
\end{center}
\end{figure}

From this figure, the temperature riches its minimum at $S_{5}^{min} \simeq 9.99$. As in the eleven Minkowski space-time associated with M-theory in the presence of $M5$ and $M2$-branes, one can show that the Hawking temperature has a minimum for
\begin{equation}
S_{5}^{min}=\frac{\pi N^{2}}{2^{3/2}},
\end{equation}
which corresponds to the minimal temperature
\begin{equation}
T_{5}^{min}=\frac{2^{3/8}}{\pi^{1/2} N^{1/4} \ell_{p}}.
\end{equation}
Under the minimal temperature, no black hole solution can exist. Above such a temperature, one can distinguish two branches. The first branch is associated with a thermodynamical unstable black hole. The second one with large entropy describes a thermodynamic stable black hole.\\
To investigate the $D3$-brane number effect on the phase transition, we exploit the computed Gibbs free energy. In figure \ref{G5N}, we plot the Gibbs free energy as a function of the Hawking temperature $T_5^{(0)}$ for different values of $N$.

The sign of the Gibbs free energy changes at the Hawking-Page temperature $T_5^{HP}$ associated with $S_5^{HP}$. From the Gibbs free energy given in Eq.\eqref{G50}, the Hawking-Page phase transition happens at
\begin{equation}
S_5^{HP}=\pi N^2,
\end{equation}
giving the following Hawking-Page temperature
\begin{equation}
T_5^{HP}=\frac{3}{2 \times 2^{1/8} \pi ^{1/2} N^{1/4} \ell_{p}},
\end{equation}
being smaller than $T_5^{min}$.
In figure \ref{G5N}, a first order Hawking-Page phase transition can occur at $T_5^{HP}$ between large (stable) black holes and the thermal radiation. Besides, we see that below the higher point representing $T_5^{min}$ no black hole can survive. For lower Gibbs free energy values, we find two branches. The upper branch describes a small (unstable) black hole with negative specific heat values. The lower branch, however, represents (large) a stable black hole solution with positive specific heat values.

\begin{figure} 
\begin{center}
\includegraphics[scale=0.55]{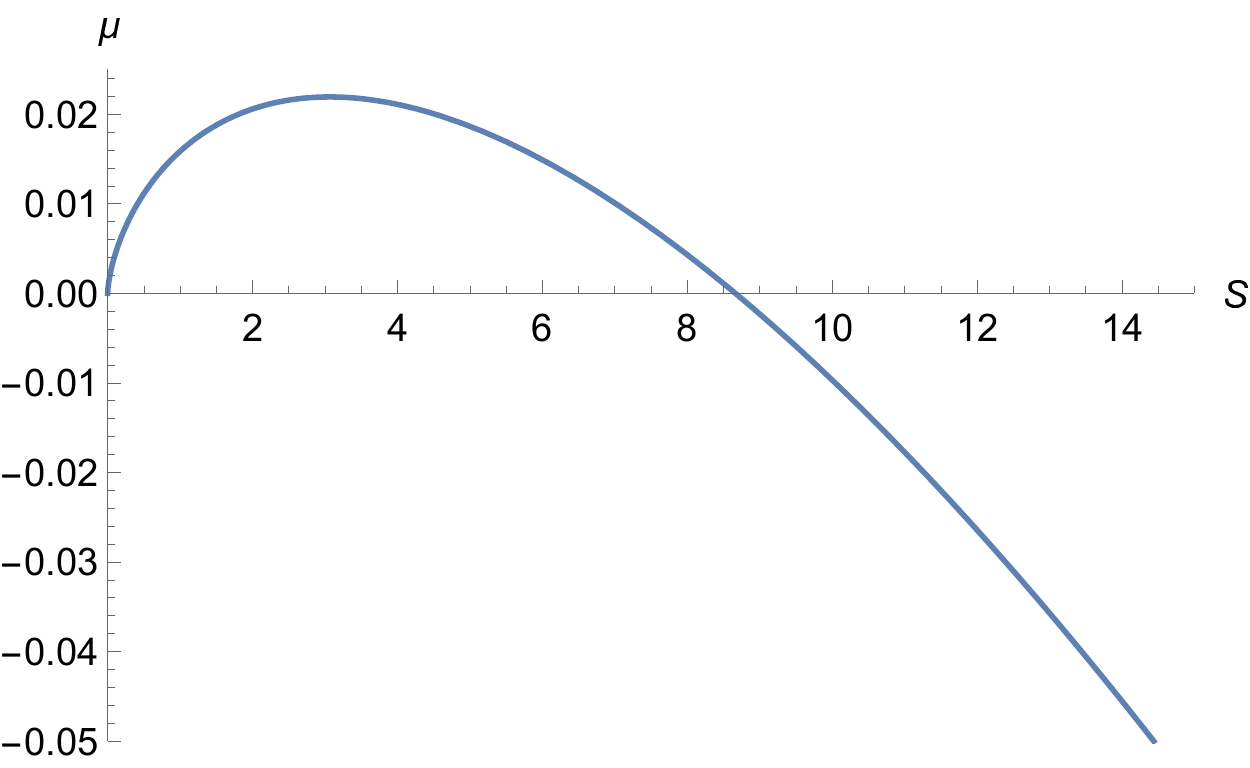}
\caption{The chemical potential as a function of entropy for $N = 3, \ell_{p}=1$. The sign change of the chemical potential happens at $S_5^{\mu=0} \simeq 8.6$.}
\label{mu5}
\end{center}
\end{figure}

The chemical potential can be also used to discuss phase transitions of the five-dimensional $AdS$ black hole in type IIB superstring. In figure \ref{mu5}, we plot the chemical potential as a function of the entropy $S$ for a fixed value of the $D3$-brane number. For simplicity reasons, we take $N=3$.

\begin{figure} 
\begin{center}
\includegraphics[scale=0.55]{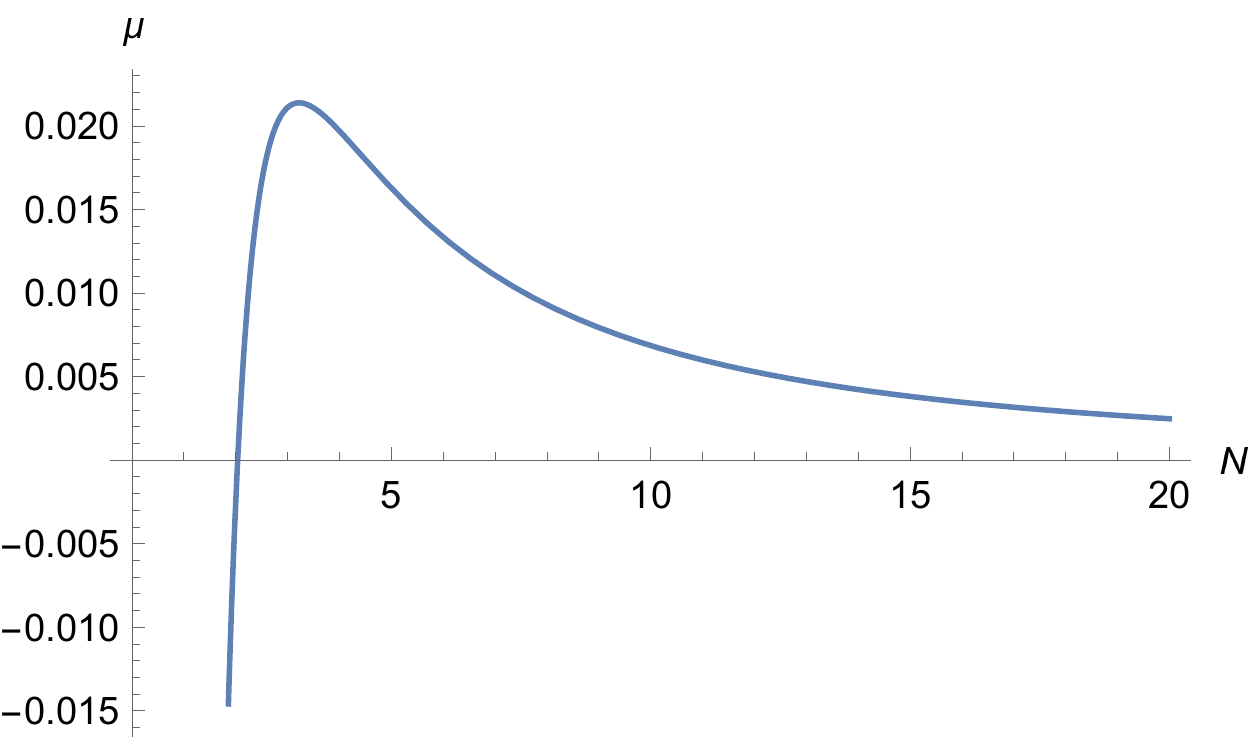}
\caption{The chemical potential as a function of the number of $D3$-branes $N$. Here we take $\ell_{p} = 1$ and $S = 4$. }
\label{mu5N}
\end{center}
\end{figure}

We notice that the chemical potential is positive for a small entropy $S$, and negative for a large
entropy $S$. The sign change, in the chemical potential, occurs at the entropy
\begin{equation}
S_5^{\mu=0}=\left( \frac{5 }{11 } \right)^{3/2} \pi N^{2}.
\label{s5mu0}
\end{equation}
Using this equation, one has $S_5^{\mu=0}<S_5^{min}<S_5^{HP}$ which is the same as the seven-dimensional $AdS$ black hole case appearing in M-theory in the presence of $M5$-branes.  We observe, then, that for the odd values of the dimension $d$ we have $S_d^{\mu=0}<S_d^{min}<S_d^{HP}$. \\
The entropy of Eq.\eqref{s5mu0} corresponds to the temperature
\begin{equation}
T_5^{\mu=0}=\frac{21}{2 \times 2^{1/8} \sqrt{55} \,  \pi^{1/2} \, \ell_p \, N^{1/4}}.
\label{t5mu0}
\end{equation}
In figure \ref{mu5N}, the chemical potential is plotted as a function of $N$ in the case of a fixed entropy $S$.

\begin{figure} 
\begin{center}
\includegraphics[scale=0.55]{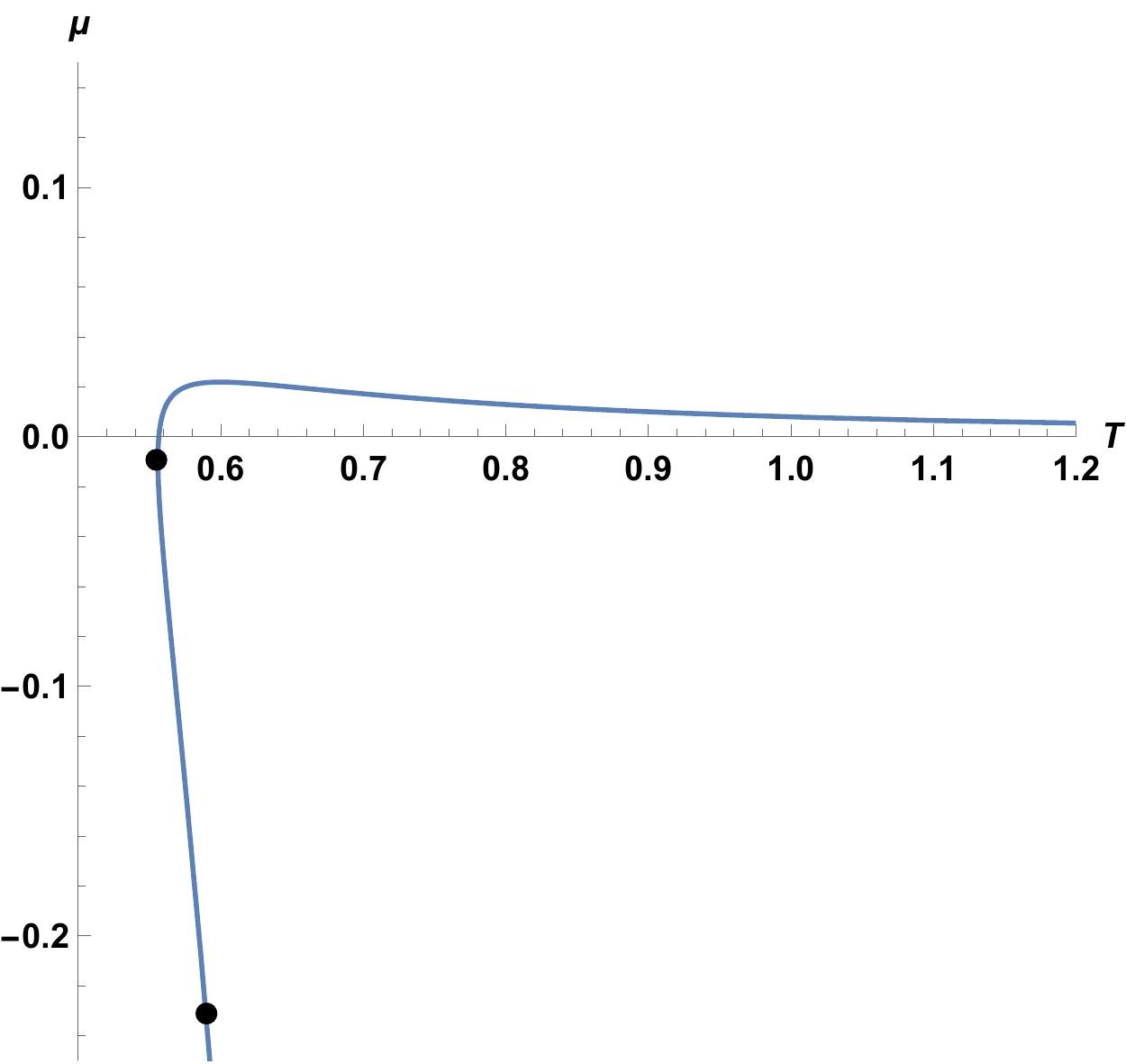}
\caption{The chemical potential as a function of the temperature $T_5^{(0)}$. We take $\ell_{p} = 1$ and $N = 3$.}
\label{mu5T}
\end{center}
\end{figure}

From figure \ref{mu5N}, the maximum of the chemical potential corresponds to the value $ S_5^{max} = \left( \frac{19 }{77 } \right)^{3/2} \pi \left( N_5^{max} \right)^{2} $, namely, $N_5^{max} \simeq 3.2 $ for $S_5^{max}=4$.\\
In figure \ref{mu5T}, we plot the chemical potential as a function of the temperature $T_5^{(0)}$ for a fixed value of $N$.

In figure \ref{mu5T}, the lower dot denotes $T_5^{HP}$ and the upper one indicates $T_5^{min}$.
The Hawking-Page temperature separates the lower stable branch and the upper unstable branch where $T_5^{min}$ resides. In this case, the chemical potential changes the sign at an unphysical region. From Eq.\eqref{s5mu0}, the chemical potential is positive when the number of $D3$-branes, $N$, is constrained by the condition $N^{2} > \frac{11^{3/2} }{5^{3/2}} \, \frac{S}{\pi}$. This limit can be saturated for
\begin{equation}
T^{\mu=0}_5 \simeq 0.9 \, T_5^{HP}.
\end{equation}
We remark that $T_5^{\mu=0} < T_5^{HP}$ which implies that the pure $AdS_5$ geometry is preferred over the black hole backgrounds.

\section{The $AdS_ 4\times \mathbb{S}^{7}$ case in presence of DE}

We consider now the model labeled by the triplet $(11,4,3)$ associated with the compactification of M-theory on 
$\mathbb{S}^7$ in the presence of $M2$-branes. Using \eqref{omegaAds4}, we obtain
\begin{itemize}
\item the mass
\begin{equation}
M_{4}^{(3) \, DE}(S,N,c)=M_{4}^{(3)}(S,N) -\alpha^{DE}_4(S,N,c),
\label{MDE40}
\end{equation}

\item the temperature
\begin{equation}
T_{4}^{(3) \, DE}(S,N,c)=T_{4}^{(3)}(S,N) +\frac{3\omega_q}{2} \cdot \frac{\alpha^{DE}_4(S,N,c)}{S},
\label{TDE40}
\end{equation}

\item the chemical potential
\begin{equation}
\mu_{4}^{(3) \, DE}(S,N,c)=\mu_{4}^{(3)}(S,N) - \frac{7 \left( 3\omega_q + 2 \right)}{18} \cdot \frac{\alpha^{DE}_4(S,N,c)}{N^{3/2}},
\label{muDE40}
\end{equation}
\item the Gibbs free energy
\begin{equation}
G_{4}^{(3) \, DE}(S,N,c)=G_{4}^{(3)}(S,N) - \left( \frac{3\omega_q+2}{2} \right)\cdot \alpha^{DE}_4(S,N,c).
\label{GDE40}
\end{equation}
\end{itemize}
In all these quantities associated with $M2$-branes, $\alpha^{DE}_4(S,N,c)$ which denotes the DE contribution reads as
\begin{equation}
\alpha^{DE}_4(S,N,c)=\frac{ c \, \pi^{\frac{5\omega_q}{4}-\frac{1}{6}} \, N^{\frac{7 \omega_q}{4} + \frac{7}{6}}}{2^{\frac{31 \omega_q}{4} + \frac{37}{6}} \, 3^{\frac{3 \omega_q +2 }{2}} \, S^{\frac{3 \omega_q}{2}} \,  \ell_p^{3\omega_q+2}}.
\label{alpha4}
\end{equation}
For such a quintessential $AdS$ black hole solution, it has been remarked that the parameter $\omega_q$ has a fixed value which is $-1/3$. As in the previous M-theory model, this can be obtained from the dimensional analysis of the thermodynamical quantities given in Eqs. \eqref{MDE40}, \eqref{TDE40}, \eqref{muDE40} and \eqref{GDE40} since they are all proportional to $\ell_{p}^{-1}$.

\begin{figure} 
\begin{center}
\includegraphics[scale=0.55]{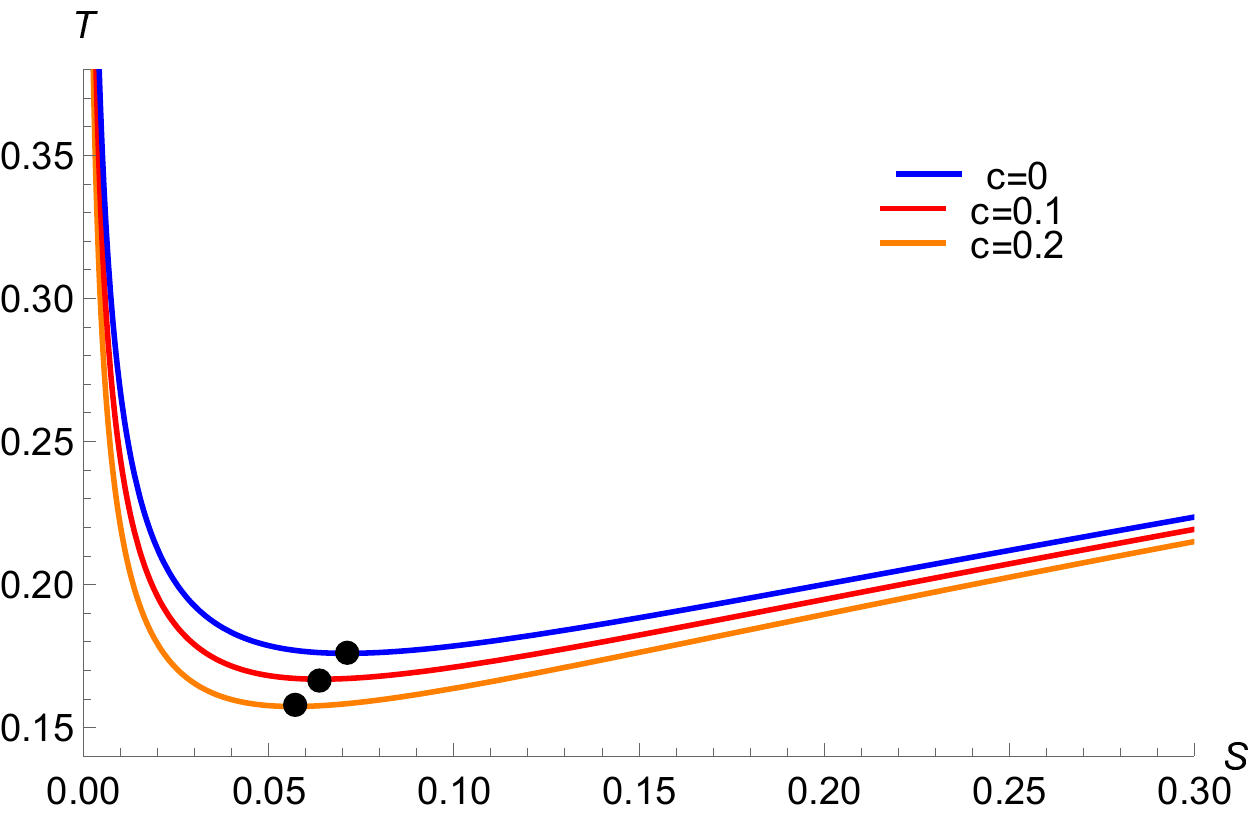}
\caption{Temperature as function of the entropy using $N=3$ and $\ell_{p}=1$ and different values of $c$.}
\label{T4DE}
\end{center}
\end{figure}

To study the phase transitions behavior in the presence of quintessence, we first plot the behavior of the Hawking temperature with respect to the entropy in figure \ref{T4DE} for different values of the intensity $c$ of the quintessence field providing DE contributions.

\begin{figure} 
\begin{center}
\includegraphics[scale=0.45]{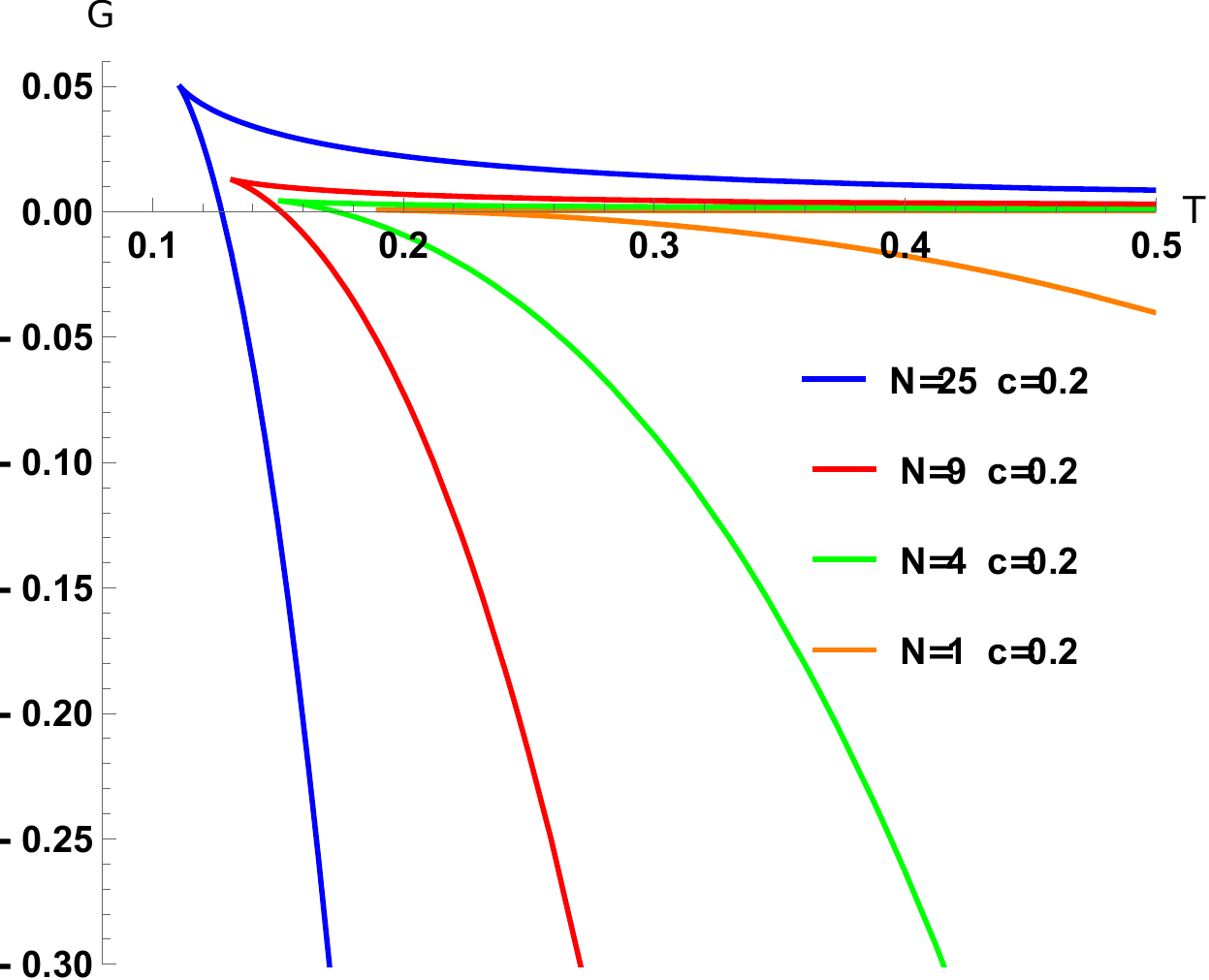}
\caption{The Gibbs free energy as function of the Hawking temperature for different values of $N$, $\ell_{p}=1$ and $c=0.2$.}
\label{G4NDE}
\end{center}
\end{figure}

From figure \ref{T4DE}, we observe that the temperature is decreasing when the intensity of DE $c$ increases. For a general number $N$ of $M2$-branes and a generic intensity $c$, the Hawking temperature has a minimum at
\begin{equation}
S_{4}^{DE-min}=\frac{\pi^{3/2}}{3^{2} \, 2^{11/2}} \left( 1-c \right)N^{3/2}.
\end{equation}
This corresponds to the following minimal temperature
\begin{equation}
T_{4}^{DE-min}=\frac{3^{1/2} \sqrt{1-c}}{2^{5/6} \, \pi^{4/3} N^{1/6} \ell_{p}}.
\end{equation}
To get more information about the phase transitions of the four-dimensional $AdS$ black hole in M-theory, the Gibbs free energy as a function of the Hawking temperature $T_4^{(3) DE}$ for different values of $N$ and a fixed value of $c$ is illustrated in figure \ref{G4NDE}.

\begin{figure} 
\begin{center}
\includegraphics[scale=0.4]{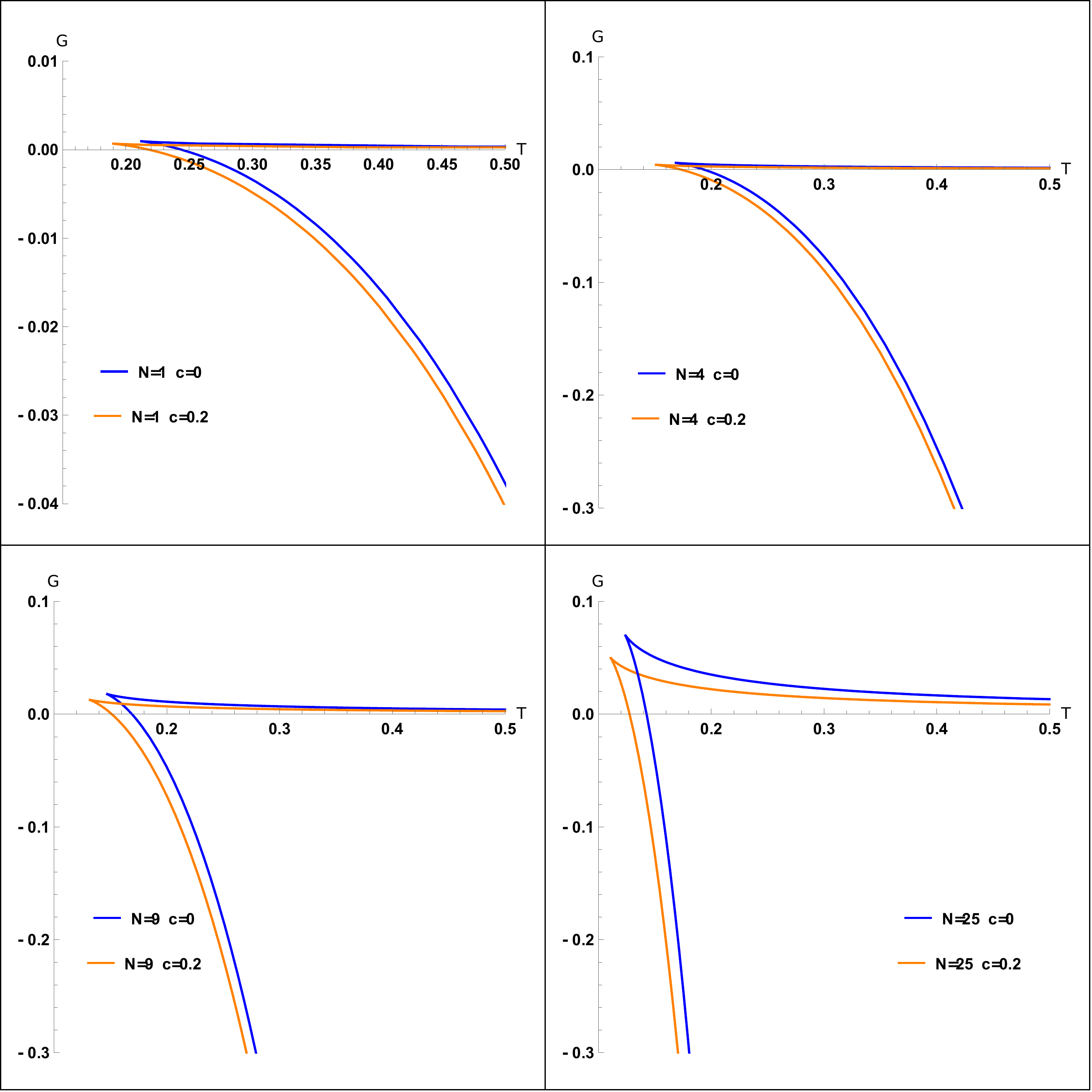}
\caption{The Gibbs free energy as function of temperature for different in the absence of DE and in its presence for different values of $N$, $\ell_{p}=1$.}
\label{G4DE}
\end{center}
\end{figure}

In figure \ref{G4NDE}, we observe that the phases are smaller than the ones discussed in figure \ref{G4N}. The Gibbs free energy decreases producing a smaller unstable black hole phase.
Using the Gibbs free energy, given in Eq.\eqref{GDE40}, one can obtain the Hawking-Page phase transition. This occurs at the following entropy
\begin{equation}
S_{4}^{DE-HP}=\frac{\left( 1-c \right) \pi^{3/2} N^{3/2}}{3 \cdot 2^{11/2}},
\end{equation}
which corresponds to the Hawking-Page temperature
\begin{equation}
T_4^{DE-HP}=\frac{2^{1/6} \sqrt{1-c}}{\pi ^{4/3} N^{1/6} \ell_{p}}.
\end{equation}
An examination shows that $T_4^{DE-HP} < T_{4}^{DE-min}$. \\
To see the difference in the Gibbs free energy behavior in the presence of quintessence, we plot this function for $c=0$ (absence of DE) and for $c=0.2$ for certain values of $N$ in figure \ref{G4DE}.

From figure \ref{G4DE}, we see that the decrease in the radiation phase and the phase where no black hole can survive comes from the   temperature diminution. However, the stable and the unstable phase are directly affected by the diminution of the Gibbs free energy.

\begin{figure} 
\begin{center}
\includegraphics[scale=0.55]{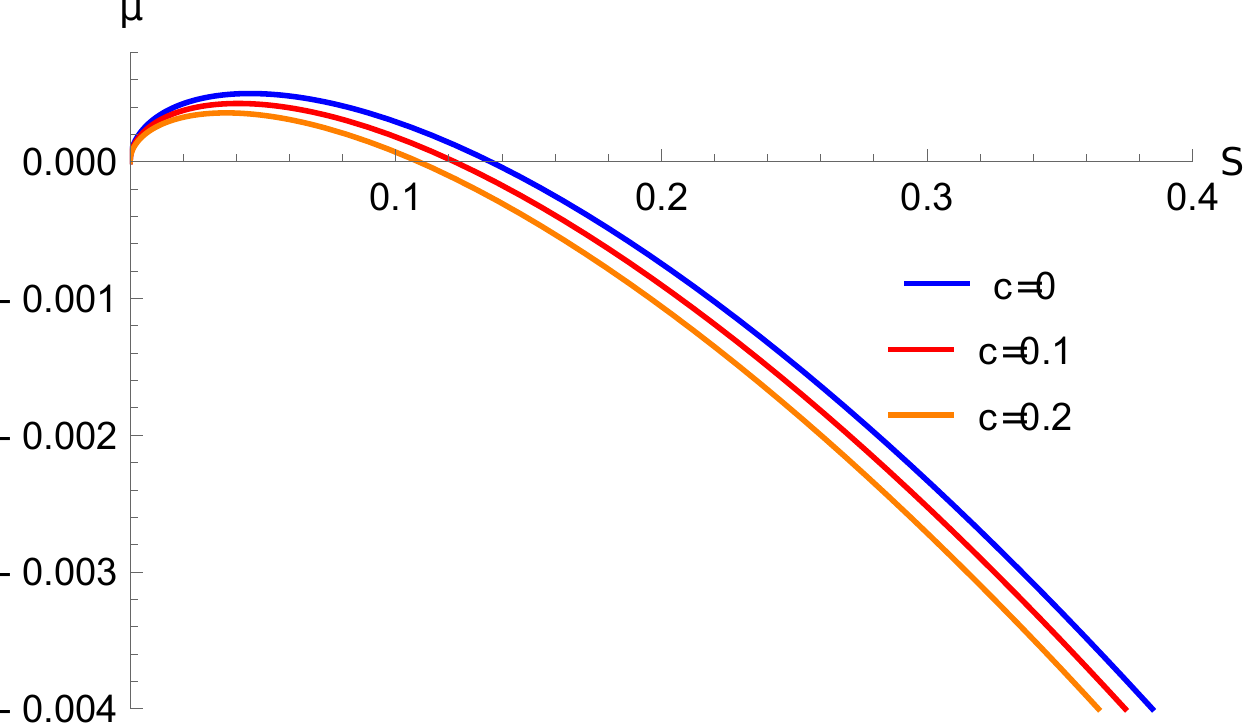}
\caption{The chemical potential as a function of entropy for $N = 3, \ell_{p}=1$.}
\label{mu4DE}
\end{center}
\end{figure}

As in the seven-dimensional $AdS$ black hole case, the quintessence field affects also the chemical potential. To examine the corresponding effects, we plot the chemical potential as a function of the entropy $S$ for $N=3$ in figure \ref{mu4DE}. It shows a diminution of the chemical potential and the entropy when DE is present.

\begin{figure} 
\begin{center}
\includegraphics[scale=0.55]{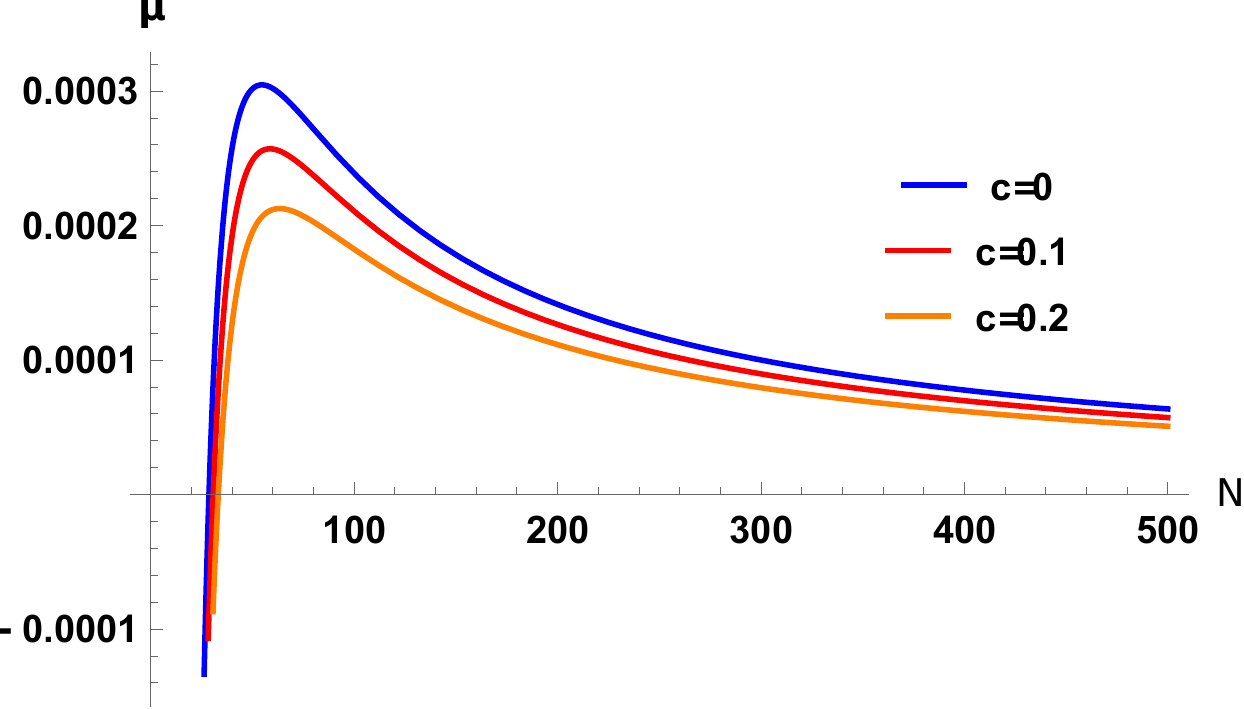}
\caption{The chemical potential as a function of the number of $M5$-branes $N$. Here we take $\ell_{p} = 1$ and $S_4^{max} = 4$. }
\label{mu4NDE}
\end{center}
\end{figure}

Moreover, we observe that the chemical potential is positive for small entropy $S$, and negative for large $S$ values. The sign change of the chemical potential happens at the entropy
\begin{equation}
S_{4}^{DE-\mu=0}=\frac{7 \, \left( 1-c \right) \pi ^{3/2} }{11 \times 3 \cdot 2^{11/2}} N^{3/2}.
\end{equation}
The four-dimensional $AdS$ black hole verifies $S_{4}^{DE-\mu=0}<S_{4}^{DE-min}<S_{4}^{DE-HP}$.\\
Furthermore, we can also study the modification of the chemical potential when it is plotted as a function of the number $N$ of $M2$-branes in M-theory. This modification is shown in figure \ref{mu4NDE}. Here, the entropy $S$ has a fixed value.

The maximum of the chemical potential corresponds to the point
\begin{equation}
 S_{4}^{DE-max} = \frac{7 \left( 1-c \right) \times \, \pi^{3/2}}{3 \times 29\, \cdot 2^{11/2}} \left( N_{4}^{DE-max} \right)^{3/2},
\end{equation}
namely, $N_{4}^{DE-max}  \simeq 63,42 $ for $S_{4}^{max}=4$ and $c=0.2$. From $N_4^{DE-max}$, we see that the number of $M2$-branes grows in the presence of DE.

\begin{figure} 
\begin{center}
\includegraphics[scale=0.55]{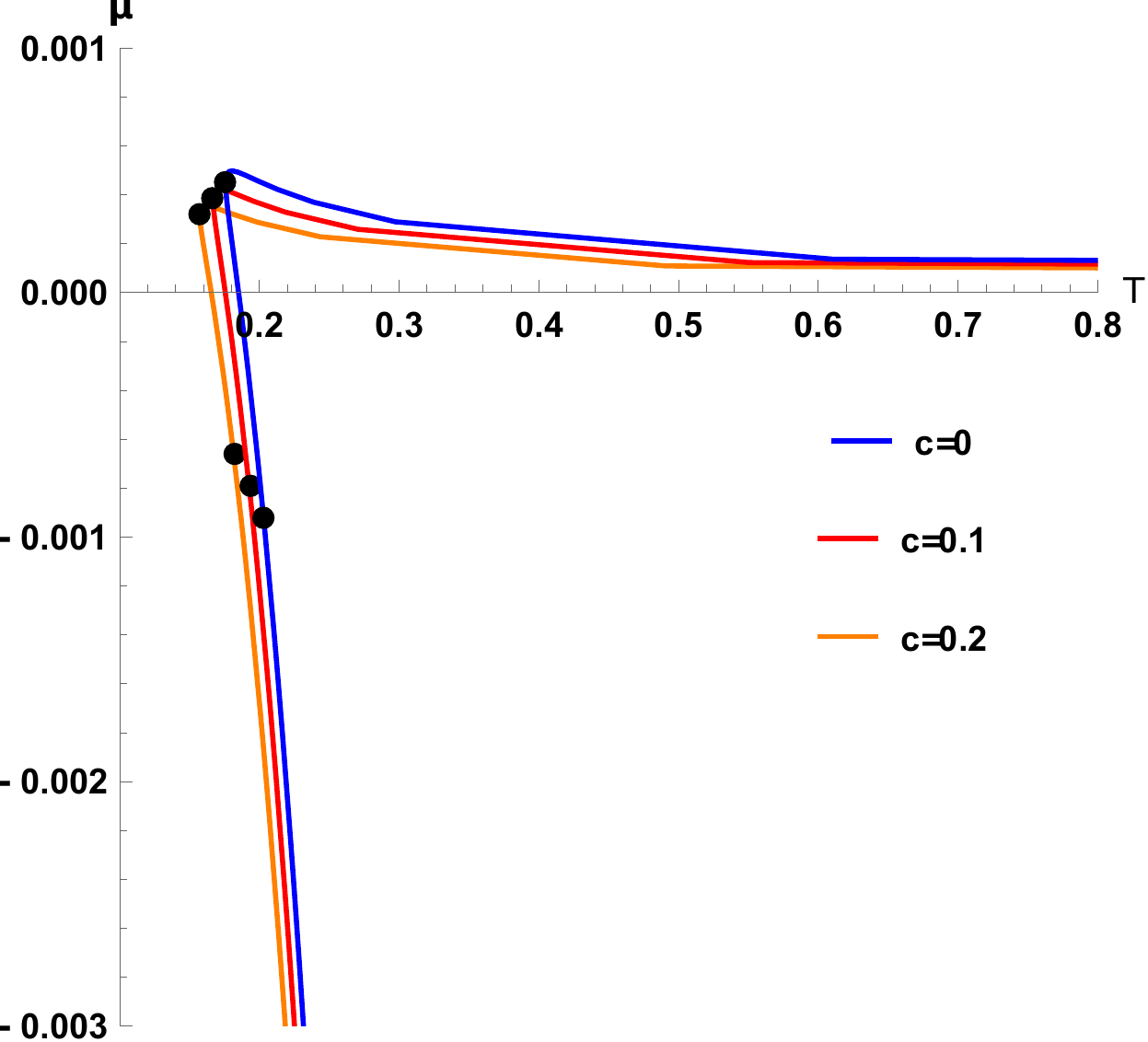}
\caption{The chemical potential as a function of temperature $T_4^{ (3) \, DE }$. We take $\ell_{p} = 1$ and $N = 3$.}
\label{mu4TDE}
\end{center}
\end{figure}

Finally, DE could be used to stabilise  the AdS black hole. To reveal this, we plot the chemical potential as a function of the temperature $T_4^{(3) \, DE}$ for a fixed number of $M2$-branes $N$  in figure \ref{mu4TDE}.

The higher dots denotes $T_4^{DE-min}$. Lower from such points, we have the $T_4^{DE-HP}$ dots which separates the lower stable branch and the upper unstable one where $T_4^{DE-min}$ resides. It is observed that, for large $c$ values, the  $T_4^{DE-min}$ is getting higher, which is also true for $T_4^{DE-HP}$. Thus, it is clear that the unstable branch is getting smaller and the stable one is bigger.

\section{The $AdS_ 5\times \mathbb{S}^{5}$ case in presence of DE}

Now, we consider a model of ten-dimensional theory associated with $D3$-brane physics. This model has been labeled by the triplet $(10,5,0)$ concerning the compactification of type IIB superstring on $\mathbb{S}^{5}$.
Using Eq.\eqref{omegaAds5}, we can obtain the  relevant quantities $X_5^{(0) \, DE}$. In particular, we list the following computations
\begin{itemize}
\item mass
\begin{equation}
M_{5}^{(0) \, DE}(S,N,c)=M_{5}^{(0)}(S,N) -\alpha^{DE}_5(S,N,c),
\label{MDE50}
\end{equation}

\item temperature
\begin{equation}
T_{5}^{(0) \, DE}(S,N,c)=T_{5}^{(0)}(S,N) +\frac{4\omega_q}{3} \cdot \frac{\alpha^{DE}_5(S,N,c)}{S},
\label{TDE50}
\end{equation}

\item chemical potential
\begin{equation}
\mu_{5}^{(0) \, DE}(S,N,c)=\mu_{5}^{(0)}(S,N) - \frac{5 \left( 4\omega_q + 3 \right)}{24} \cdot \frac{\alpha^{DE}_5(S,N,c)}{N^{2}},
\label{muDE50}
\end{equation}
\item Gibbs free energy
\begin{equation}
G_{5}^{(0) \, DE}(S,N,c)=G_{5}^{(0)}(S,N) - \left( \frac{4\omega_q+3}{3} \right)\cdot \alpha^{DE}_5(S,N,c).
\label{GDE50}
\end{equation}
\end{itemize}
In this way, the DE manifestly term $\alpha^{DE}_5(S,N,c)$ reads
\begin{equation}
\alpha^{DE}_5(S,N,c)=\frac{ 3 c \, \pi^{\frac{10\omega_q}{3}+\frac{3}{2}} \, N^{\frac{5 \omega_q}{3} + \frac{5}{4}}}{2^{\frac{ \omega_q}{2} + \frac{19}{8}}  \, S^{\frac{4 \omega_q}{3}} \, \ell_p^{4\omega_q+3}}.
\label{alpha5}
\end{equation}
For such a black hole solution in type IIB superstring compactification on $\mathbb{S}^{5}$, it has been remarked that the parameter $\omega_q$ has a fixed value being  $-1/2 $.

\begin{figure} 
\begin{center}
\includegraphics[scale=0.55]{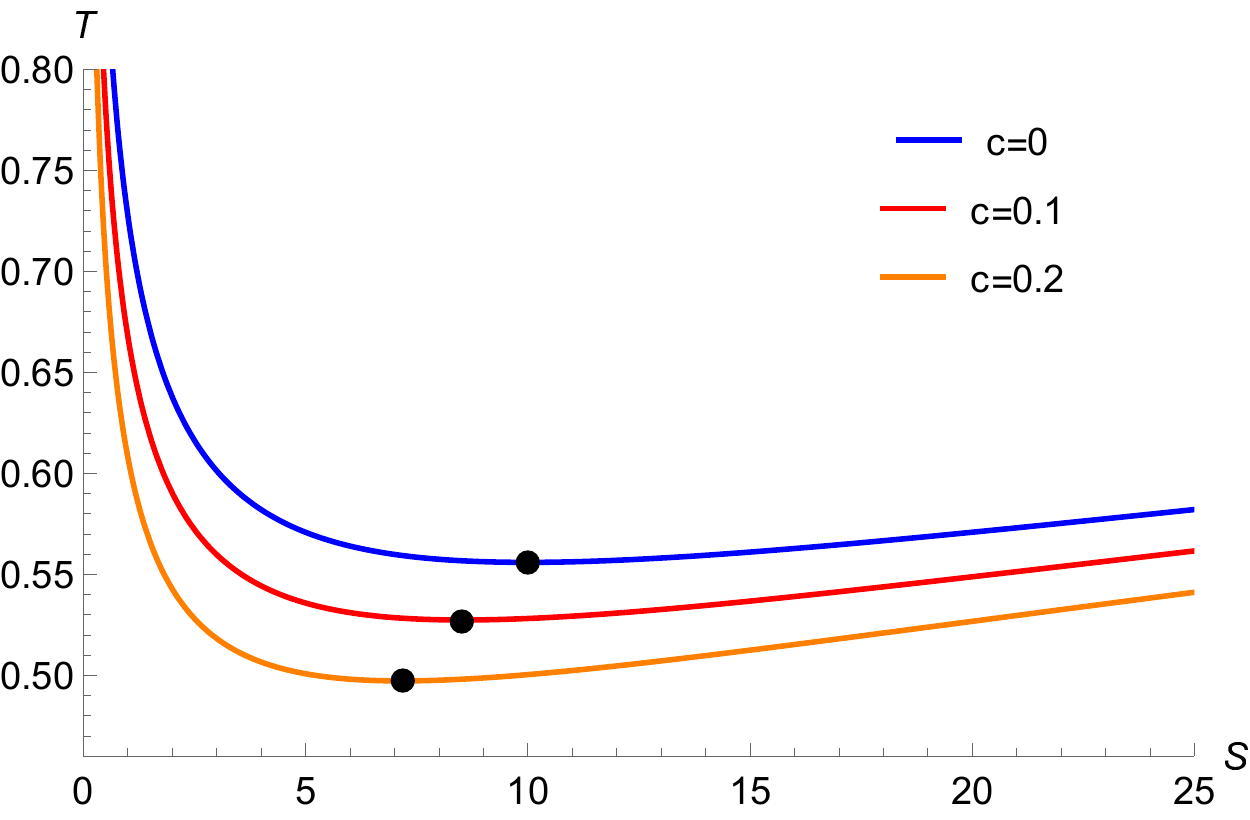}
\caption{Temperature as function of the entropy using $N=3$ and $\ell_{p}=1$ and different values of $c$.}
\label{T5DE}
\end{center}
\end{figure}

To investigate the phase transition behaviours in type IIB superstring, we first plot the Hawking temperature as a function of the entropy in figure \ref{T5DE} for different values of the intensity $c$.

The dots in figure \ref{T5DE} denote the minimum of the temperature. For a general number of $D3$-branes and a generic intensity of DE field, we can show that the Hawking temperature has a minimum at the following entropy
\begin{equation}
S_{5}^{DE-min}=\frac{\pi \left(1-c \right)^{3/2} N^{2}}{2^{3/2}},
\end{equation}
corresponding to the minimal temperature
\begin{equation}
T_{5}^{DE-min}=\frac{2^{3/8} \sqrt{1-c} }{\pi^{1/2} N^{1/4} \ell_{p}}.
\end{equation}
From figure \ref{T5DE}, we see that the temperature is decreasing when the intensity of DE $c$ increases. This confirms that DE can be considered as a cooling system surrounding the black hole \cite{notrea}. The minimum of the temperature and the corresponding entropy denoted by the black dotes in figure \ref{T5DE} are also affected in the same way.

\begin{figure} 
\begin{center}
\includegraphics[scale=0.45]{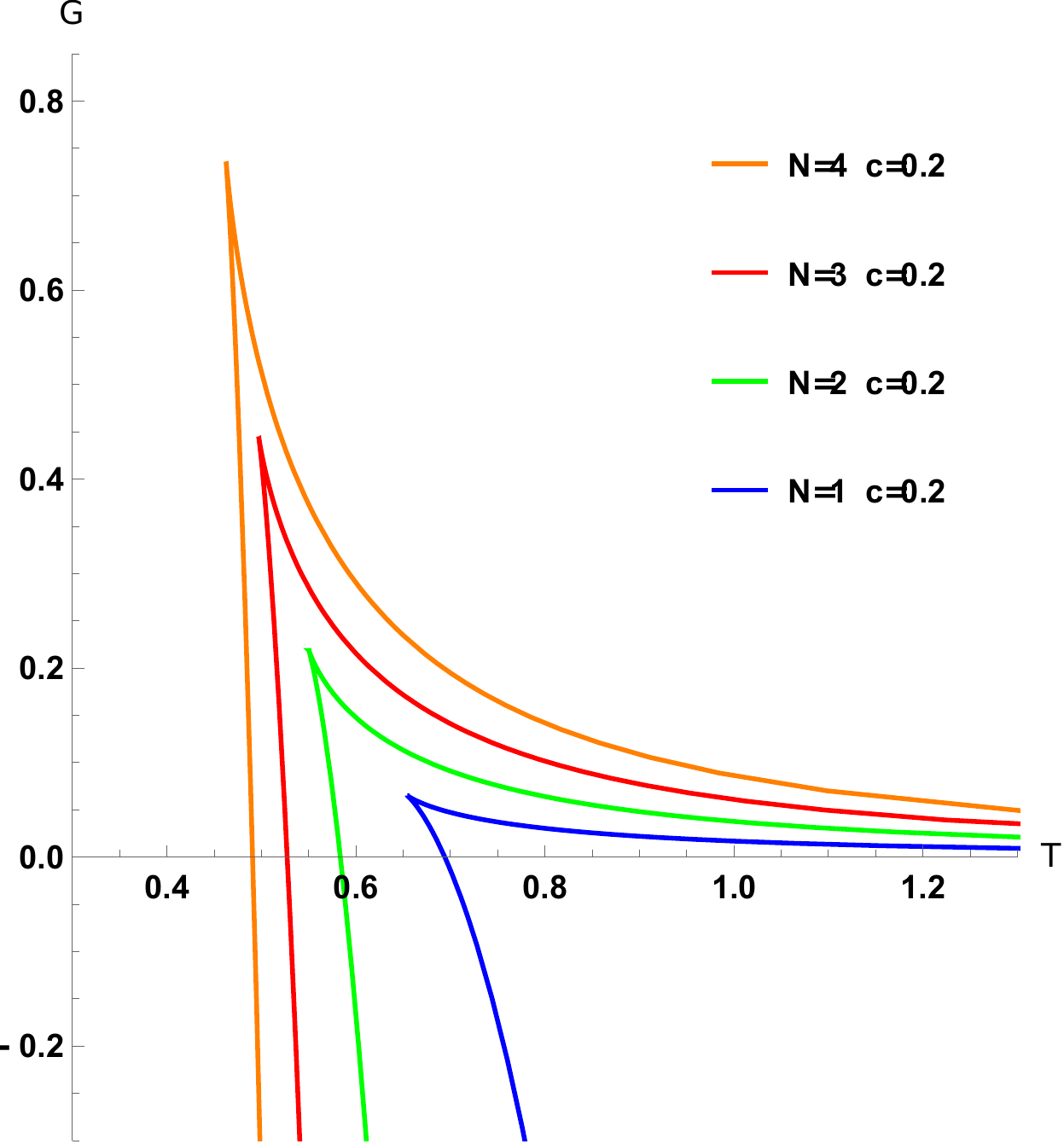}
\caption{The Gibbs free energy as function of the temperature for different values of $N$, $\ell_{p}=1$ and $c=0.2$. The sign of Gibbs free energy changes at  $S_5^{HP}$ and corresponds to the Hawking-Page temperature $T_5^{DE-HP}$.}
\label{G5NDE}
\end{center}
\end{figure}

For more investigations about the phase transitions of the five-dimensional $AdS$ black hole in type IIB superstring,  we illustrate the Gibbs free energy as a function of the Hawking temperature $T_5^{(0) \, DE}$ for certain values of the $D3$-brane number $N$ and fixed values of $c$ in figure \ref{G5NDE}.

We notice that the phases discussed previously are reduced. Indeed, the Gibbs free energy decreases generating smaller unstable black hole phase.
To see the difference in the behavior of the Gibbs free energy when the quintessence is present, we plot this function for $c=0$ (absence of DE) and $c=0.2$ for several values of $N$ associated with $D3$-branes.

\begin{figure} 
\begin{center}
\includegraphics[scale=0.4]{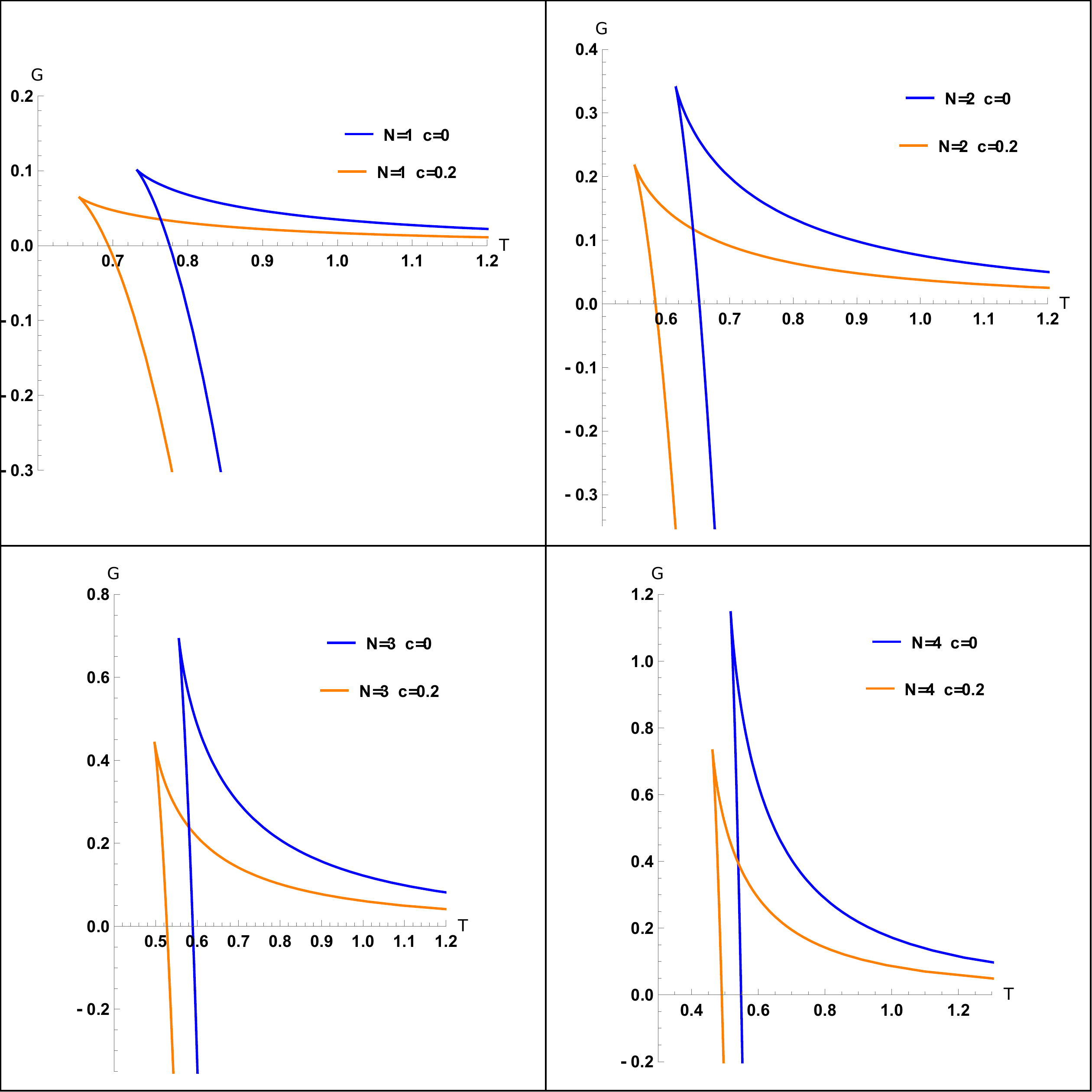}
\caption{The Gibbs free energy as function of temperature for different in the absence of DE and in its presence for different values of $N$, $\ell_{p}=1$.}
\label{G5DE}
\end{center}
\end{figure}

\begin{figure} 
\begin{center}
\includegraphics[scale=0.55]{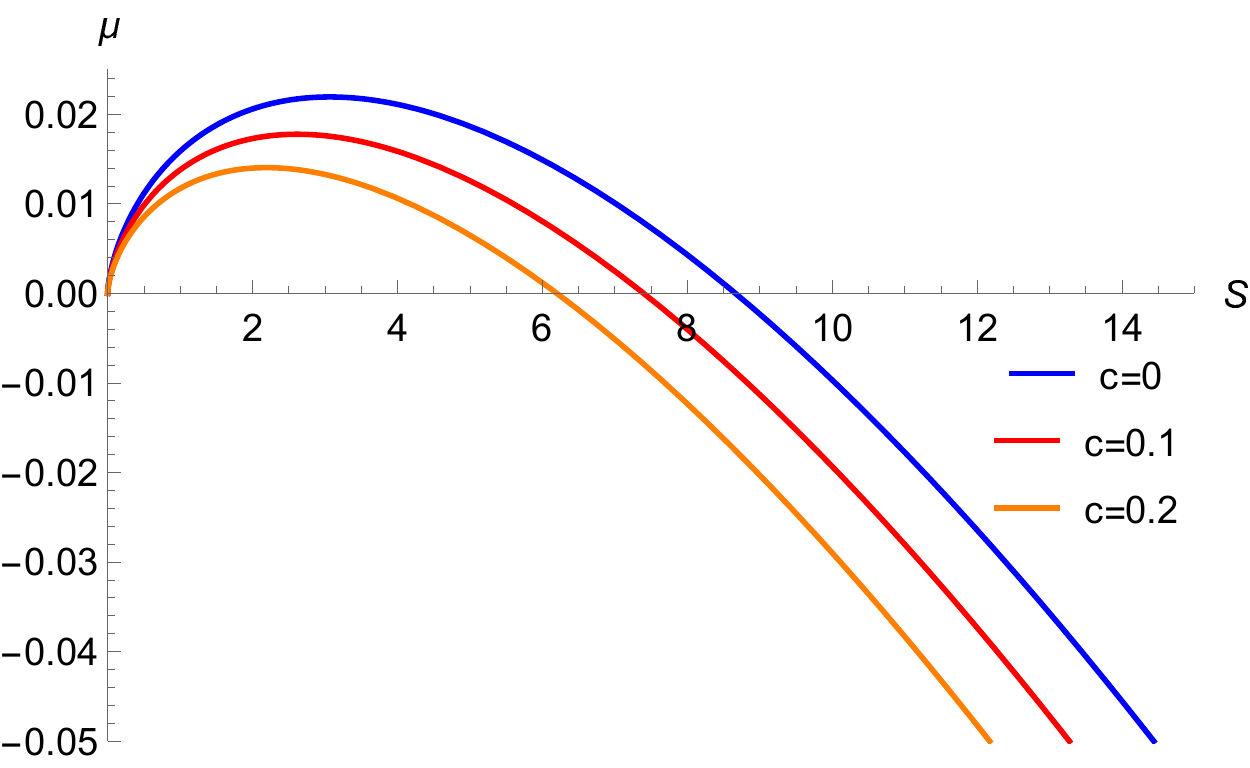}
\caption{The chemical potential as a function of entropy for $N = 3, \ell_{p}=1$. The sign change of the chemical potential happens at $S_5^{DE-\mu=0}$ .}
\label{mu5DE}
\end{center}
\end{figure}

It follows from figure \ref{G5DE} that the decrease in the radiation phase and the phase where no black hole can exist can be understood from the diminution of temperature. However, the stable and the unstable phase are directly affected by the diminution of the Gibbs free energy. From such an energy given in Eq.\eqref{GDE50}, we find the Hawking-Page phase transition which occurs at
\begin{equation}
S_5^{DE-HP}=\left(1-c \right)^{3/2} \pi N^2,
\end{equation}
corresponding to the Hawking-Page temperature
\begin{equation}
T_5^{DE-HP}=\frac{3 \, \sqrt{1-c} }{2 \times 2^{1/8} \pi ^{1/2} N^{1/4} \ell_{p}}.
\end{equation}
This temperature is smaller than $T_5^{DE-min}$.\\
As in the M-theory compactification on $\mathbb{S}^{4}$ and $\mathbb{S}^{7}$, the quintessence field affects also the chemical potential. To examine the corresponding modifications, we plot the chemical potential as a function of entropy $S$ for $N=3$ in figure \ref{mu5DE}.

\begin{figure} 
\begin{center}
\includegraphics[scale=0.55]{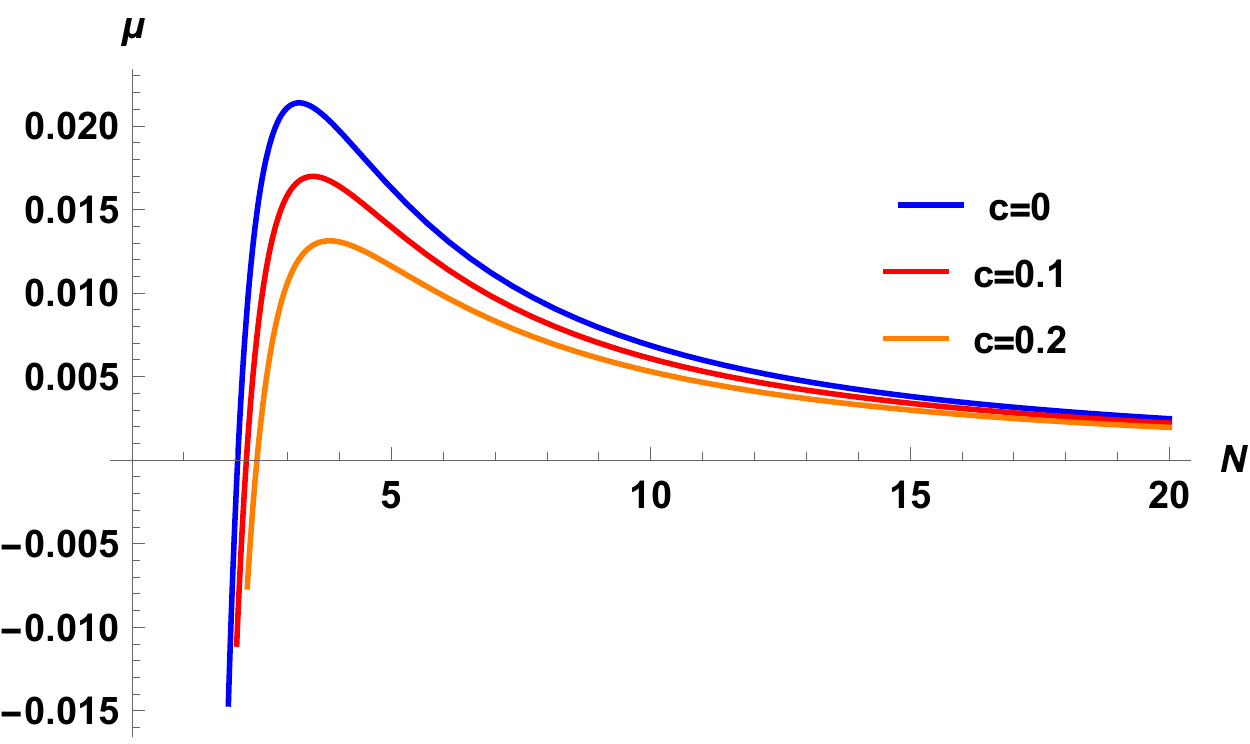}
\caption{The chemical potential as a function of the number of $M5$-branes $N$. Here we take $\ell_{p} = 1$ and $S_5^{DE-max} = 4$. }
\label{mu5NDE}
\end{center}
\end{figure}

This figure shows a diminution of both the chemical potential and the entropy when DE is present. The sign change in the chemical potential occurs at the following entropy
\begin{equation}
S_5^{DE-\mu=0}=\left( \frac{5 }{11 } \right)^{3/2}\, \left(1-c \right)^{3/2} \,  \pi N^{2}.
\end{equation}
Using this equation, one can show that $S_5^{DE-\mu=0}<S_5^{DE-min}<S_5^{DE-HP}$.\\
Furthermore, we can also discuss the chemical potential as a function of the number $N$ of $D3$-branes in type IIB superstring which is illustrated in figure \ref{mu5NDE}.

\begin{figure} 
\begin{center}
\includegraphics[scale=0.55]{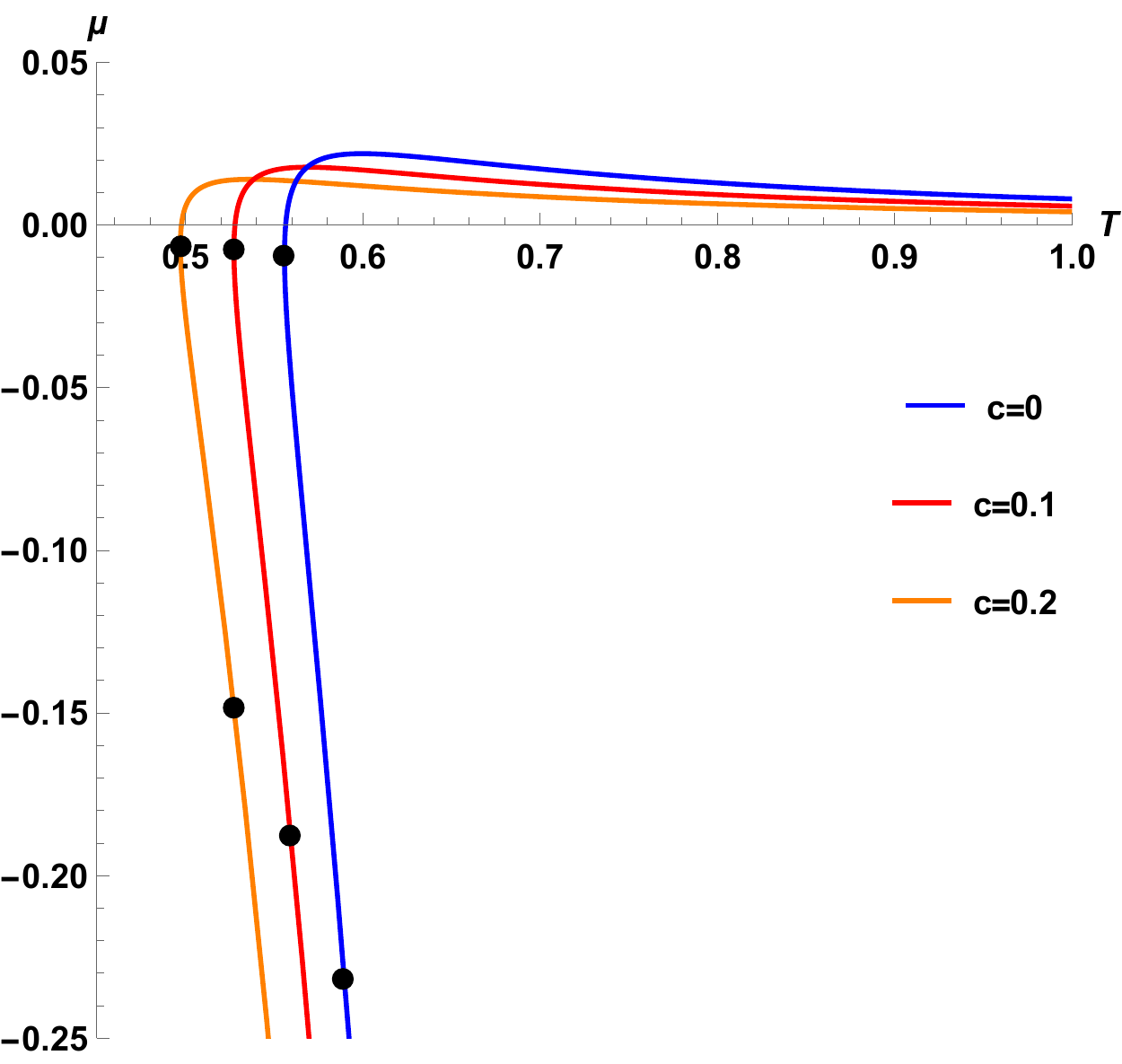}
\caption{The chemical potential as a function of temperature $T$. We take $\ell_{p} = 1$ and $N = 3$.}
\label{mu5TDE}
\end{center}
\end{figure}

The maximum of the chemical potential corresponds to the point
\begin{equation}
S_5^{DE-max} = \left( \frac{19 }{77 } \right)^{3/2} \, \left(1-c \right)^{3/2} \, \pi  \left( N_5^{DE-max} \right)^{2}
\end{equation}
namely, $N_5^{DE-max} \simeq 3.8$ for $S_5^{DE-max}=4$ and $c=0.2$. From $N_5^{DE-max}$, we see that the number $N$ of $D3$-branes grows in the presence of DE.
In figure \ref{mu5TDE}, we plot the chemical potential as a function of the temperature $T_5^{(0) \, DE}$ for a fixed number $N$ of $D3$-brane in the compactification of type IIB superstring on $\mathbb{S}^{5}$.

The higher dots  are associated with $T_5^{DE-min}$. Lower from these
 points, we have the $T_5^{DE-HP}$ point that separates the lower stable
 branch and the upper unstable branch where $T_5^{DE-min}$ resides. We
 notice that $T_5^{DE-min}$ is getting higher in the curves when the
 intensity of DE $c$ is bigger, which is also true for $T_5^{DE-HP}$. 
Thus, we observe that DE can be considered as a establishing factor for the five-dimensional quintessential $AdS$ black hole embedded in type IIB superstring theory.

\end{document}